\documentclass[aps,pre,reprint,showpacs,showkeys,amsmath,amssymb]{revtex4-1}
\usepackage{graphicx}
\newcommand{\SO}[1]{\text{SO}(#1)}
\newcommand{\St}{\text{St}}
\newcommand{\USt}{\text{USt}}

\newtheorem{definition}{Definition}[section]
\allowdisplaybreaks

\begin{document}

\title{Beyond the Rosenfeld Functional: Loop Contributions in Fundamental
Measure Theory}
\author{Stephan Korden}
\email[]{stephan.korden@rwth-aachen.de}
\affiliation{Institute of Technical Thermodynamics, RWTH Aachen
University, Schinkelstra\ss e 8, 52062 Aachen, Germany}
\date{\today}

\begin{abstract}
The Rosenfeld functional provides excellent results for the prediction of
the fluid phase of hard convex particle systems but fails beyond the freezing
point. The reason for this limitation is the neglect of orientational and
distance correlations beyond the particle diameter. In the current article we
resolve this restriction and generalize the fundamental measure theory to an
expansion in intersection centers. It is shown that the intersection probability
of particle systems is described by an algebra, represented by Rosenfeld's
weight functions. For subdiagrams of intersection networks we derive vertex
functions that provide the building blocks for the free energy functional. Their
application is illustrated by deriving the Rosenfeld functional and its leading
correction which is exact in the third virial order. Furthermore, the methods
are used to derive an approximate functional for the infinite sum over Mayer
ring diagrams. Comparing this result to the White Bear mark II functional, we
find general agreement between both results.
\end{abstract}
\pacs{61.20.Gy, 64.10.+h, 61.30.Cz}
\keywords{fundamental measure theory, resummation, vertex function}
\maketitle
\section{Introduction}\label{sec:introduction}
Historically, density functional theory (DFT) for classical particles has been
investigated in parallel to its corresponding application in quantum mechanics
\cite{evans-rev}, although with less impact on our understanding of the
underlying physics. One reason is a practical one, as Newton's equation is
much more accessible by numerical methods than Schr\"odinger's. The expensive
and difficult problem of constructing a suitable functional is therefore more
profitable in the quantum case as for classical systems \cite{qm-dft}.
Nevertheless, the analytical form of a free energy functional contains rich
information about the physical system that would otherwise be difficult to
obtain from computer experiments alone. The construction of a classical density
functional is therefore of great interest from a theoretical point of view. 

An important step forward was the development of the fundamental measure theory
(FMT) for hard convex particles by Rosenfeld \cite{rosenfeld-structure,
rosenfeld-freezing, rosenfeld-closure, rosenfeld-mixture, rosenfeld1},
generalizing the semi-heuristic scaled-particle theory of Reiss, Frisch, and
Lebowitz \cite{scaled-particle-1}. Starting from the observation that Mayer's
$f$ function is decomposable into a pairwise convolute of weight functions
\cite{isihara-orig, kihara-1, kihara-2}, the free energy functional of
uncorrelated particles is the sum of three contributions, constrained by the
scaling behavior of the weight functions. This functional and its corrections
by Rosenfeld and Tarazona \cite{tarazona, comparision-ros} proves to be of
surprising accuracy. Compared to computer simulations, the phase diagrams for
spheres, cylinders, rods, and their mixtures are in excellent agreement in the
fluid region \cite{schmidt-dft, schmid-c, schmidt-mixtures, schmidt-colloidal,
schmidt-one-dim, schmidt-palelet} and the direct vicinity of the freezing point
\cite{goos-mecke}. On the other hand, the functional fails for higher particle
densities. The reason for this shortcoming is the missing correlation between
orientations and distances beyond the two-particle system. This restriction
reflects the underlying Percus-Yevick approximation which breaks down for
highly correlated particle configurations such as crystalline structures.

Several important approaches have been made to analyze and improve the Rosenfeld
functional. A central step in this direction is the geometrically motivated
correction term introduced by Tarazona \cite{tarazona}. Whereas a different
approach compares simulation data to the structure of the functional, resulting
in the White Bear functionals \cite{white-bear-1, white-bear-2}. A different
strategy was followed by Leithall and Schmidt \cite{schmidt-diagram} by
introducing a diagrammatic formulation relating the functional to a degeneration
of Mayer diagrams.

In a recent article, we started to investigate and clarify the mathematical
origin of the local splitting of Mayer's $f$ function \cite{korden-1, korden-2}.
First it was shown that the kinematic formula of integral geometry, developed
by Blaschke, Santalo, and Chern \cite{blaschke, chern-1, chern-2, chern-3,
santalo-book} corresponds to Rosenfeld's decomposition of the second virial
integrand. More generally, the intersection probability of any number of
particles, with a common intersection center, is determined by the Euler form
and factorizes into a convolute of local densities. This result not only allowed
the derivation of Rosenfeld's functional from first principles and without
reference to the semi-heuristic scaled-particle theory, but also related the
approach to Mayer's virial expansion \cite{mcdonald}. 

In the current article, this connection between FMT and the virial expansion
will be further investigated. It will be shown how to derive higher order terms
of the free energy functional, extending the methods of \cite{korden-2}. As a
first example, the leading order correction will be obtained, which resolves the
angular degeneration of the Rosenfeld functional and clarifies how to correct
the distance correlation beyond one particle diameter and therefore exceed the
Percus-Yevick approximation.

In order to keep the article self-contained, the algebraic results of
\cite{korden-2} are repeated and refined in section \ref{subsec:review} and
generalized to an intersection algebra in Sec.~\ref{subsec:inter_diagrams}.
Deriving its representation in vertex functions in Sec.~\ref{subsec:vertex}, we
calculate the leading correction to Rosenfeld's functional in
Sec.~\ref{subsec:functional} and finally compare the result to the White Bear
mark II functional.
\section{Review of intersection networks and their Euler forms}
\label{subsec:review}
The recent investigation \cite{korden-2} has not only shown how to derive
Rosenfeld's functional from the virial expansion, but also proposed a natural
generalization to higher order corrections, where the methods developed for the
leading order will also be of relevance for all further correction terms. We
will therefore first give a short summary of the most relevant results obtained
so far, including the graphical representation of intersection networks, the
algebraic decoupling of the Euler form into weight functions, and their
resummation into a generating function.

The understanding of hard particle physics begins with the observation that the
intersection probability of particles, overlapping in at least one common point,
around which the particles can freely rotate and translate, is determined by the
Euler form. This central result of integral geometry has been derived for
two-particle domains by Blaschke, Santalo, and Chern \cite{blaschke,
santalo-book, chern-1, chern-2, chern-3} and further extended to an arbitrary
number of particles in \cite{korden-1}. The decoupling of the second virial
integral into Rosenfeld's weight functions is therefore only a specific example
of a more general relationship between differential geometry and the local
Euclidean group $\text{ISO}(3)$ of rotations and translations. 

With the exception of the second virial integral, the Euler form does not
determine Mayer integrals exactly. However, completely connected Mayer clusters
can be approximated by the intersection probability of particles that intersect
in at least one common point. The Euler form determines therefore an essential
part of these important Mayer integrals, but at the same calculational costs as
the second virial integral itself. Nevertheless, completely contracted diagrams
are only the leading order in an expansion of the free energy functional in the
number of intersection centers, which constitute the smallest unit of an
intersection network. Because of their importance, we introduced the name
``stack'' and ``universal stack'' in \cite{korden-2}, defined by:
\begin{equation}\label{stack}
\St_k = \bigcap_{i=1}^k D_i\quad,\qquad \USt = \bigoplus_{k=2}^\infty \St_k
\end{equation}
for the $i=1,\ldots, k$ particle domains $D_i$ intersecting in at least one
common point. 

The second virial integrand is equivalent to Mayer's $f$ function. This
guarantees an exact relationship between Mayer diagrams and their representation
as intersection networks of only pairwise overlapping particles. For the
simplest cluster diagrams these can be illustrated as 2-dimensional drawings.
However, to simplify the graphical representation, we also introduced
``intersection diagrams'' in \cite{korden-2}, where particles are reduced to
lines and intersection centers indicated by edge joints. 
\begin{figure}
\includegraphics[width=3.1cm,angle=-90]{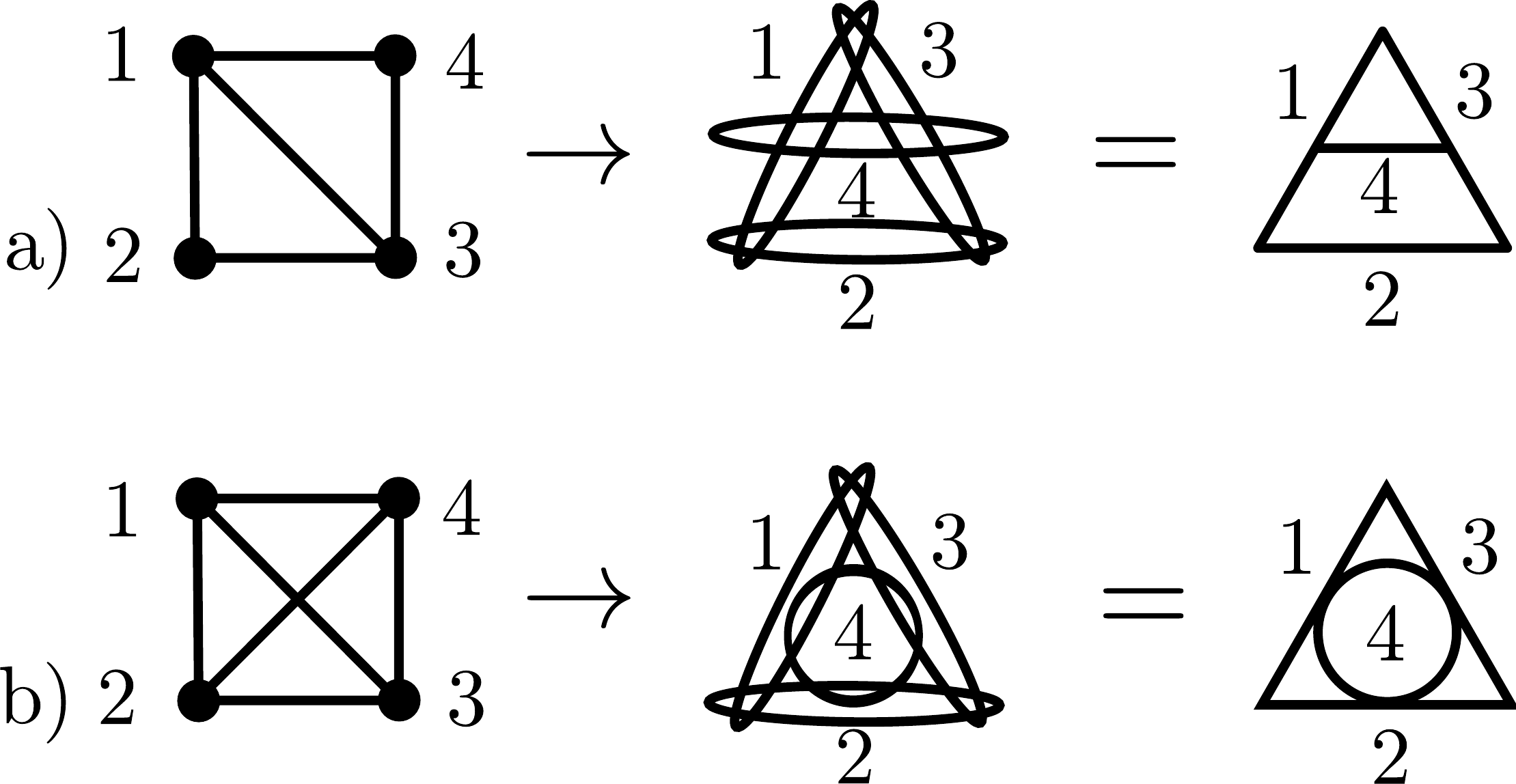}
\caption{Mayer clusters and intersecting diagrams provide identical
representations for pairwise intersecting particles, with nodes and edges
interchanged. This is shown for the two diagrams in the Mayer (left), particle
(middle), and intersection (right) representation.}
\label{fig:fourth-virial}
\end{figure}
An example with all three different types of representations is shown in
Fig.~\ref{fig:fourth-virial}. Approximations of these diagrams are derived by
the successive contraction of intersection centers
\begin{figure}
\includegraphics[width=3.9cm,angle=-90]{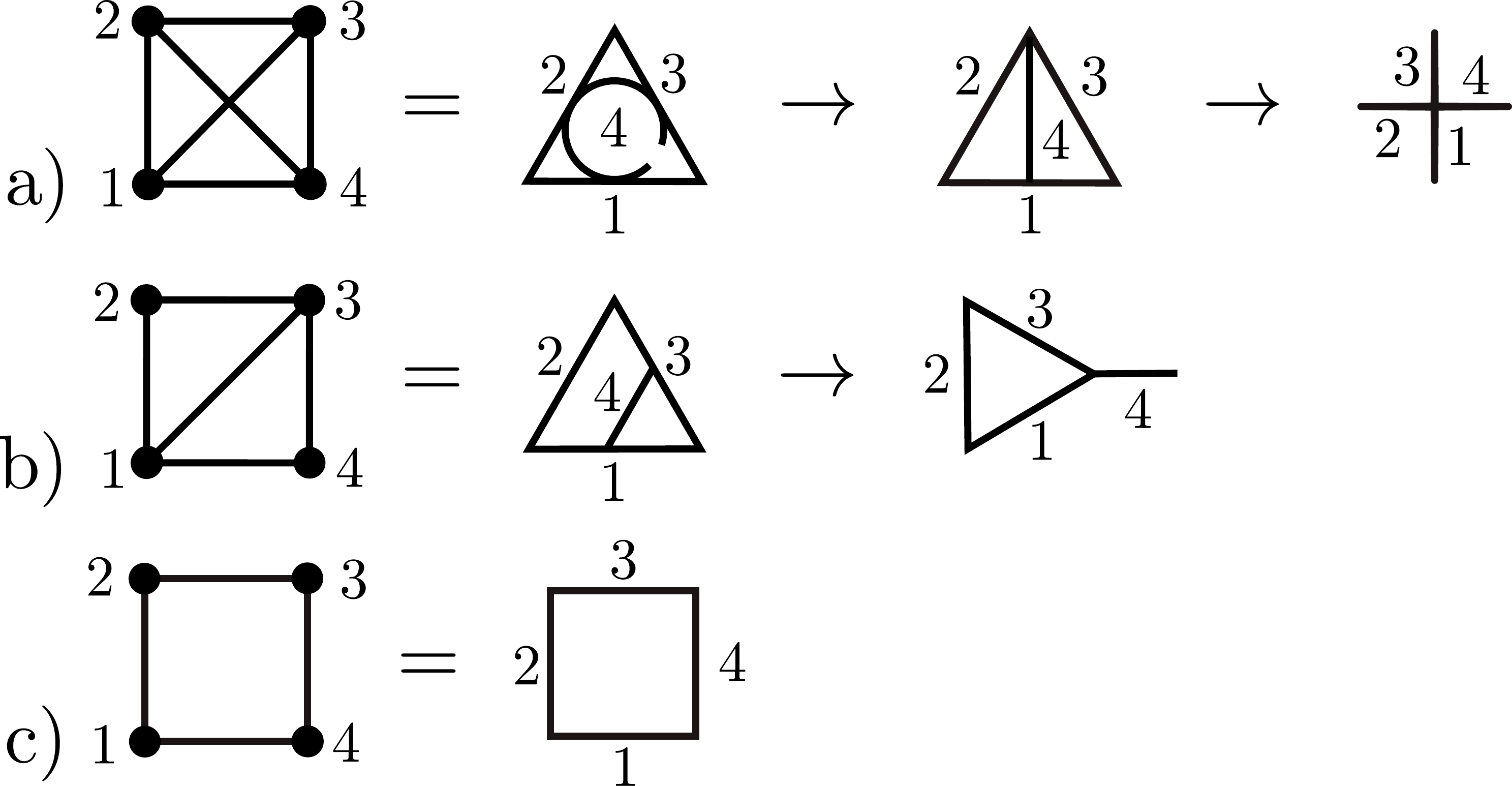}
\caption{Integrals of intersection diagrams are approximated by consecutive
contraction of their intersection centers. The example shows the three
4-particle Mayer clusters, their dual intersection diagrams, and their allowed
contractions defined by the Mayer clusters.}
\label{fig:four-virial}
\end{figure}
as shown in Fig.~\ref{fig:four-virial} for the four particle cluster diagrams.

Intersection diagrams can be classified by their number of intersection centers
$h$ and internal loops $g$. Taking this into account, the excess free-energy
functional density is the infinite sum
\begin{equation}\label{expansion}
\Phi = \sum_{g=0, h=1}^\infty \Phi_{g,h}\;,
\end{equation}
where each element $\Phi_{g,h}$ corresponds itself to an infinite set of
diagrams. 
\begin{figure}
\includegraphics[width=1.2cm,angle=-90]{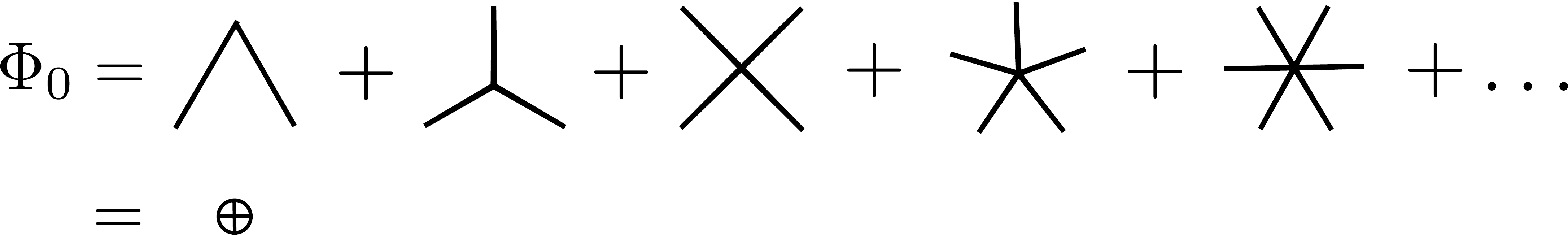}
\caption{The starfish like diagrams follow from the completely connected Mayer
clusters by maximally contraction of their intersection centers. Resummation of
all these diagrams is indicated by the crossed circle which corresponds to the
Rosenfeld functional.}
\label{fig:zero-loop}
\end{figure}
The leading element, $\Phi_{0,1}$, is presented in Fig.~\ref{fig:zero-loop} and
provides the graphical representation of Rosenfeld's functional as the
intersection probability of the universal stack. Observe that similar diagrams
have been used in \cite{schmidt-diagram} to relate the virial expansion to FMT.

The intersection probability for each stack $\St_k$ is determined by the Euler
form $K(\partial \St_k)$, integrated over the intersection domain $\vec r_a \in
\St_k$ and averaged over the positions and rotations of each individual particle
\begin{equation}
\begin{split}
\Gamma(D) := \{\, \gamma = (\vec r &, \vec \Omega)\, |\, \vec r \in D,\, \vec
\Omega \in \SO{3}\, \}\\[0.4em]
d\gamma_i & := d^3r_i\,  d^3\Omega_i\;.
\end{split}
\end{equation}
To keep the notation for the coordinates of particles and their intersection
centers apart, their indices will be labeled by the characters:
\begin{equation}\label{index_coord}
\begin{array}{ll}
a, b, c, \ldots & \text{: indices of intersection domains}\\
i,\; j,\; k,\; \ldots & \text{: indices of particle domains\;.}
\end{array}
\end{equation}

Suitably normalized \cite{korden-2} and multiplied by the single particle
densities $\rho_i$, the Euler form determines the virial coefficient at 0-loop
order
\begin{equation}\label{beta}
\begin{split}
\beta_{k-1}^{(0,1)} &=
\frac{1}{4\pi} \int_{\St_k\times \Gamma(D_2\times \ldots \times D_k)}
\hspace{-1em}K(\partial\St_{k-1})\,\delta(\vec n\vec r_a)\,d^3r_a\\[0.4em]
&\qquad\quad  \times \rho_1(\gamma_1)\ldots \rho_k(\gamma_k)\,
d\gamma_2\wedge \ldots \wedge d\gamma_k\;,
\end{split}
\end{equation}
taking into account the symmetry factor $\sigma_{k-1} = 1$ of the fully
connected Mayer diagrams. The delta-function $\delta(\vec n\vec r_a)$ of the
scalar-product of the normal and position vector restricts the integrand to the
surface $\partial D$ and follows the calculational rules introduced in Appendix
B of \cite{korden-2}.

The Euler form is a linear functional on the boundary of manifolds, which
vanishes for any odd-dimensional domain. The derivation of $K(\partial \St_k)$
for 3-dimensional particles thus reduces to only three contributions.
Introducing the surface $\Sigma$ of a domain $D$ and the notations: 
\begin{equation}
\Sigma = \partial D\;,\quad \Sigma^n = \underbrace{\Sigma\,\cap\, \ldots\,
\cap\, \Sigma}_{n-\text{times}}\,,
\end{equation}
the boundary of a stack of $k$ identical particles reduces to
\begin{equation}\label{boundary_stack}
\begin{split}
\partial \St_k & = k\, \Sigma\cap \St_{k-1} + k (k-1)\, \Sigma^2\cap \St_{k-2}\\
& \quad + k(k-1)(k-2)\, \Sigma^3 \cap \St_{k-3}
\end{split}
\end{equation}
as the intersection probability between the set of points $\Sigma^3$ and a
further surface element $K(\Sigma^3\cap\Sigma)=0$ is zero. The only
non-vanishing terms of the Euler form are therefore:
\begin{equation}\label{splitting_1}
\begin{split}
K(\Sigma)\;\, & = \,\omega_\chi\\
K(\Sigma^2) & = C^{\alpha_1\alpha_2}\omega_{\alpha_1}\omega_{\alpha_2}\\
K(\Sigma^3) & = C^{\alpha_1\alpha_2\alpha_3}\omega_{\alpha_1}
\omega_{\alpha_2}\omega_{\alpha_3}\;,
\end{split}
\end{equation}
with the $C$-matrices $C^{\alpha_1\alpha_2}$, $C^{\alpha_1\alpha_2\alpha_3}$
explicitly derived in \cite{korden-2}. As first shown by Chern \cite{chern-2},
these tensors are independent of the particle geometry and solely defined by the
Euler form and the dimensions of the particles and their embedding space. 

To this list of algebraic relations it is useful to add a further one, which
follows from the tensorial density of the integrand
(\ref{beta}):
\begin{equation}\label{splitting_2}
K(\Sigma^n\cap \St_k) = K(\Sigma^n)\, (\omega_v)^k\;.
\end{equation}

For 3-dimensional convex particles it has been shown in \cite{korden-2} that the
infinite dimensional basis set of weight functions can be grouped into five
classes:
\begin{align}\label{weight-functions}
\omega_\chi(\vec r_a -\vec r_i)&=\frac{1}{4\pi}\kappa_G\delta(\vec n\vec r_a)
\nonumber\\
\omega_{\kappa L}(\vec r_a -\vec r_i)&=\frac{1}{4\pi} \bar\kappa
\vec{n}^{\otimes L}\delta(\vec n \vec r_a) \nonumber\\
\omega_{\Delta L}(\vec r_a -\vec r_i)&=\frac{1}{4\pi} \Delta \vec{n}^{\otimes
L} 
\delta(\vec n \vec r_a)\\[0.4em]
\omega_{\sigma L}(\vec r_a-\vec r_i)&=\vec{n}^{\otimes L} \delta(\vec n 
\vec r_a)
\nonumber \\[0.5em]
\omega_v(\vec r_a -\vec r_i)&=\Theta(\vec r_a - \vec r_i)\;,\nonumber
\end{align}
corresponding to the Euler-characteristic $\chi$ of the Gauss curvature
$\kappa_G$, the mean curvature $\bar \kappa$, the curvature difference or
tangential curvature $\Delta$, the surface $\sigma$, and the particle volume
$v$. Each of these geometric terms is taken at the intersection center $\vec
r_a$ with respect to their absolute position in the embedding space $\vec r_i$.
As a consequence of the non-algebraic splitting of the scalar Euler form, the
weight functions also depend on the $L$-fold tensor product of the normal vector
$\vec n$ for $L~\in~\mathbb{N}_0$.

The weight function $\omega_v$ of the particle volume $v$ plays a special role.
It is not part of the curvature dependent Euler form (\ref{splitting_1}) but
constrains the integration domain in (\ref{beta}) from the embedding space
$\mathbb{R}^3$ to the particle volume. The relation (\ref{splitting_2}) is
therefore a formal one. Nonetheless, it is useful to include it to the set of
algebraic relations (\ref{splitting_1}) and to introduce two different indices
for the weight functions
\begin{equation}\label{index_weigth}
\begin{split}
A, B,C\ldots &\in\{v,\chi,\kappa L,\sigma L\}\\
\alpha, \beta, \gamma \ldots &\in\{\chi,\kappa L,\sigma L\} \;,
\end{split}
\end{equation}
indicating if $v$ is included or not. The characters are chosen such that to
each intersection center $\vec r_a\in \St_k$ the family of indices
$(a,A_1,\alpha_1,A_2,\alpha_2,\ldots)$, $(b,B_1,\beta_1,\ldots)$,
$(c,C_1,\gamma_1,\ldots)$, \ldots is assigned, providing an
intuitive relation between weight indices and intersection points.

It is a special property of the 0-loop order (\ref{beta}) and its single
intersection center that each particle domain $D_i$ is related to a single
weight function $\omega_A^i$. Only in this special case, it is possible to
combine them into the 1-point density
\begin{equation}\label{dens-weight}
n_A(\vec{r}_a) = \sum_{i=1}^M \int_{\Gamma(D_i)}
\rho_i(\gamma_i)\,\omega^i_A(\vec{r}_a-\vec{r}_i) d\gamma_i
\end{equation}
introduced by Rosenfeld \cite{rosenfeld-structure}. Whereas higher loop orders
require the definition of the ``$k$-point function''
\begin{equation}\label{k-point-func}
\prod_{p=1}^k \omega_{A_p}^i(\vec r_{a_p}-\vec r_i)
\end{equation}
for $k$ disjunct intersection centers $\vec r_{a_p}$ at particle domain
$\vec r_i\in D_i$ and its corresponding ``$k$-point density``
\begin{equation}\label{k-point}
\begin{split}
&n_{A_1,\ldots, A_k}(\vec r_{a_1},\ldots, \vec r_{a_k})\\
& \quad = \sum_{i=1}^M \int_{\Gamma(D_i)} \rho_i(\gamma_i)\,
\prod_{p=1}^k \omega_{A_p}^i(\vec r_{a_p}-\vec r_i)\,d\gamma_i\;,
\end{split}
\end{equation}
as introduced in \cite{korden-2}, generalizing Eq.~(\ref{dens-weight}) and
Wertheim's 2-point measure \cite{wertheim-1, wertheim-2,wertheim-3,
wertheim-4}. 

Due to the coupling of the particle density to the weight functions, the free
energy $F$ is no longer a functional of $\rho_i$ alone. Instead, $F$ now depends
on the new variable $n_v$. As has been shown in \cite{korden-2}, the weight
function $\omega_v$ acts as the neutral element under removing $\St_k \to
\St_{k-1}$ or adding $\St_k \to \St_{k+1}$ a particle and thus shifting the
Euler form (\ref{splitting_2}) by one factor of $\omega_v$. The corresponding
shifts for the boundary of the stack (\ref{boundary_stack}) or its Euler form
\begin{equation}\label{euler-stack}
\begin{split}
K(\partial \St_k) & = k\,K(\Sigma)\,\omega_v^{k-1} + k(k-1)\,
K(\Sigma^2)\,\omega_v^{k-2}\\
&\quad + k(k-1)(k-2)\,K(\Sigma^3)\,\omega_v^{k-3}
\end{split}
\end{equation}
are generated by integration and differentiation with respect to $\omega_v$:
\begin{equation}\label{int-diff}
\begin{split}
\int K(\partial \St_k)\,d\omega_v & = \frac{1}{k+1}\, K(\partial \St_{k+1})\\
\frac{\delta K(\partial \St_k)}{\delta \omega_v} &= k\, K(\partial
\St_{k-1})
\end{split}
\end{equation}
and thus relate intersection integrals (\ref{beta}) for more than three
particles $k\geq 3$. These operations allow to translate the virial expansion in
the particle density representation $\rho$ to that in the weight density $n_v$.
To see this, consider the virial expansion of the chemical potential
\cite{mcdonald}, written for constant $\rho$ and $\beta^{-1}=\text{k}_B T$:
\begin{equation}\label{virial-integral}
\begin{split}
\beta_{k-1}&  = \frac{1}{V}\frac{\sigma_k}{(k-1)!} \int f_{1,2}\ldots
f_{k-1, k} \,d\gamma_1 \ldots d\gamma_k\\
&\quad \beta\mu = \beta \mu_{\text{id}} + \sum_{k=2}^\infty \beta_{k-1}
\rho^{k-1}\;,
\end{split}
\end{equation}
depending on the embedding volume $V$ and the symmetry coefficient $\sigma_k$.

From this derives the free energy potential by adding one further particle,
realized as the integral over $\rho$:
\begin{equation}\label{virial_old}
\begin{split}
\beta F & = \beta F_{\text{id}}+ \beta F_{\text{ex}}
= \int \beta \mu(\rho)\, d\rho\\
& \;\, \beta F_{\text{ex}} = \sum_{k=2}^\infty \; \int \frac{1}{k}\;
\beta_{k-1}\rho^k\;.
\end{split}
\end{equation}

In the representation of weight densities, the last step corresponds to the shift
$\St_{k-1}\to \St_k$ in the particle stack, which for $k\geq 3$ particles is
realized as the integral over $n_v$. 

The operations (\ref{int-diff}) therefore apply to the free energy functional,
defined as the generating function of cluster integrals. With the functional
derivative
\begin{equation}
\frac{\delta n_A (\vec{r}_a)}{\delta n_B ( \vec{r}_b)} 
= \delta_{AB}\,\delta(\vec r_a - \vec r_b)\;,
\end{equation}
the free energy $F$ is related to the free energy density $\Phi$ by the integral
\begin{align}\label{func-def}
\beta F & = \int \frac{\delta (\beta F)}{\delta n_v(\vec r_a)}
\delta n_v(\vec r_a) = \int \frac{\delta (\beta F)}{\delta n_v(\vec r_a)}\, 
dn_v(\vec r_a)\,d^3r_a \nonumber\\
& =: \int \Phi(\vec r_a) \, d^3r_a \nonumber\\
& = \sum_{g,h} \int \Phi_{g,h}(\vec r_{a_1},\ldots \vec r_{a_h})
\,d^3r_{a_1}\ldots d^3r_{a_h}\,,
\end{align}
which reduces to the expansion in intersection diagrams by
Eq.~(\ref{expansion}). Comparing this result to the virial representation of the
free energy (\ref{virial_old}) yields a relation between the density functionals
$\Phi_{g,h}$ and their corresponding cluster densities
\begin{equation}\label{virial}
\begin{split}
\Phi([n_A], \vec r_a) & = \int \sum_{k=2}^\infty \; \frac{1}{k}\; 
\beta_{k-1}([n_A],\vec r_a)\, dn_v\\
& = \int \sum_{k=2}^\infty \; \frac{1}{k}\; 
\sum_{g,h}\;\beta_{k-1}^{(g,h)}([n_A],\vec r_a)\, dn_v
\end{split}
\end{equation}
that is uniquely defined up to an integration constant. Its value has been
determined in \cite{korden-2} and corresponds to the formal definition of a
single particle virial coefficient
\begin{equation}
\beta_0 = \beta_0^{(0,1)} := n_\chi\;.
\end{equation}

The final step in proving the equivalence of the intersection probability of
the universal stack and Rosenfeld's functional consists in deriving the first
element $\Phi_{0,1}$ of (\ref{func-def}). Inserting the algebraic
representations (\ref{splitting_1}), (\ref{splitting_2}) into the Euler form
(\ref{euler-stack}) and rewriting the cluster integral (\ref{beta}) in 1-point
densities
\begin{align}\label{beta-1}
& \frac{1}{k}\beta^{(0,1)}_{k-1}([n_A]) = \int \Bigl[ \omega_\chi^{i_1}
(\omega_v^{k-1})^{i_2\ldots i_k} \Bigr.\nonumber\\
& + (k-1)\,C^{\alpha_1\alpha_2} \omega_{\alpha_1}^{i_1} \omega_{\alpha_2}^{i_2}
(\omega_v^{k-2})^{i_3\ldots i_k}\nonumber\\[0.4em]
& \Bigl.\; + (k-1)(k-2)\, C^{\alpha_1\alpha_2 \alpha_3} \omega_{\alpha_1}^{i_1}
\omega_{\alpha_2}^{i_2} \omega_{\alpha_3}^{i_3} (\omega_v^{k-3})^{i_4\ldots
i_k}\Bigr] \nonumber\\[0.4em]
& \qquad \times \rho_{i_1} \ldots \rho_{i_k}\, d\gamma_{i_2}\ldots d\gamma_{i_k}
\, d^3r_a\\[0.4em]
& = \int \Bigl[ n_\chi n_v^{k-1} + 
(k-1)\,C^{\alpha_1\alpha_2} n_{\alpha_1} n_{\alpha_2} n_v^{k-2}\Bigr.
\nonumber\\
&\quad \Bigl.  +\, (k-1)(k-2)\,C^{\alpha_1\alpha_2\alpha_3} n_{\alpha_1}
n_{\alpha_2} n_{\alpha_3} n_v^{k-3}\Bigr] \, d^3r_a\nonumber
\end{align}
yields the decoupled integral for a stack of order $k$. Adding up all cluster
integrals and integrating over the packing density $n_v$
\begin{align}\label{ros-resum}
&\Phi_{0,1}([n_A], \vec r_a) = \int \sum_{k=1}^\infty\frac{1}{k}
\beta_{k-1}^{(0,1)}([n_A],\vec r_a)\, dn_v(\vec r_a)\\
&= \int \left[ \frac{n_\chi}{1-n_v} 
+ C^{\alpha_1\alpha_2} \frac{n_{\alpha_1} n_{\alpha_2}}{(1-n_v)^2}
\right.\nonumber\\
& \left. \qquad \qquad\qquad +\, 2\,C^{\alpha_1\alpha_2\alpha_3}
\frac{n_{\alpha_1} n_{\alpha_2} n_{\alpha_3}}{(1-n_v)^3}\;\right]\; 
dn_v(\vec r_a)\nonumber
\end{align}
reproduces Rosenfeld's functional
\begin{align}\label{ros-2}
\Phi_{0,1}&= -n_\chi\,\ln{(1-n_v)}\\[0.4em]
&\quad + C^{\alpha_1\alpha_2} \frac{n_{\alpha_1}
n_{\alpha_2}}{1-n_v} + C^{\alpha_1\alpha_2\alpha_3} \frac{n_{\alpha_1}
n_{\alpha_2} n_{\alpha_3}}{(1-n_v)^2}\nonumber
\end{align}
and identifies it as the 0-loop order of the expansion in intersection centers. 

The simple structure of this result is explained by the single intersection
point, as shown in Fig.~\ref{fig:zero-loop}. However, higher order intersection
diagrams, as shown in Fig.~\ref{fig:four-virial}, not only incorporate further
intersection points but also loop constraints that create corrections to the
direct correlation function at distances larger than the particle diameter.
Going beyond the Percus-Yevick approximation requires therefore the introduction
of additional mathematical tools.
\section{The intersection algebra}\label{subsec:inter_diagrams}
Intersection diagrams were introduced in \cite{korden-2} to visualize the
approximation scheme of FMT and to associate the virial expansion of the free
energy to the Euler form and thus to Rosenfeld's functional. They also gave a
first intuitive understanding of higher order corrections as partially
contracted diagrams. Figures Fig.~\ref{fig:fourth-virial} and
Fig.~\ref{fig:four-virial} demonstrate how they can be obtained by graphical
construction. For more complex diagrams, however, this approach becomes unwieldy
and algebraic rules for their construction and manipulation are more useful.

As an example, we will first reconsider the intersection diagrams of
Fig.~\ref{fig:fourth-virial} and Fig.~\ref{fig:four-virial} in detail in the
first paragraph \ref{subsubsec:contraction_rules} and then generalize the
results to diagrams of arbitrary degree of contraction in
\ref{subsubsec:algebra}.
\subsection{Intersection diagrams}
\label{subsubsec:intersection_diagrams}
The Euler form provides a unique identity between Mayer's $f$ function and the
weight functions
\begin{equation}\label{f-function}
\begin{split}
& f_{ij}(\vec r_i - \vec r_j) \\
& = \int_{D_i\cap D_j} C^{A_1A_2}\omega_{A_1}^i(\vec r_a - \vec r_i)
\omega_{A_2}^j(\vec r_a - \vec r_j)\, d^3r_a
\end{split}
\end{equation}
and relates the representation in particle $\vec r_i, \vec r_j$ and intersection
coordinates $\vec r_a$. Thus any sequence of $f$ functions $f_{12} f_{13} f_{14}
\ldots$, multiplied by their particle densities, can uniquely be rewritten in
$k$-point densities (\ref{k-point}) as a function of their intersection
coordinates.

\begin{figure}
\includegraphics[width=3.5cm,angle=-90]{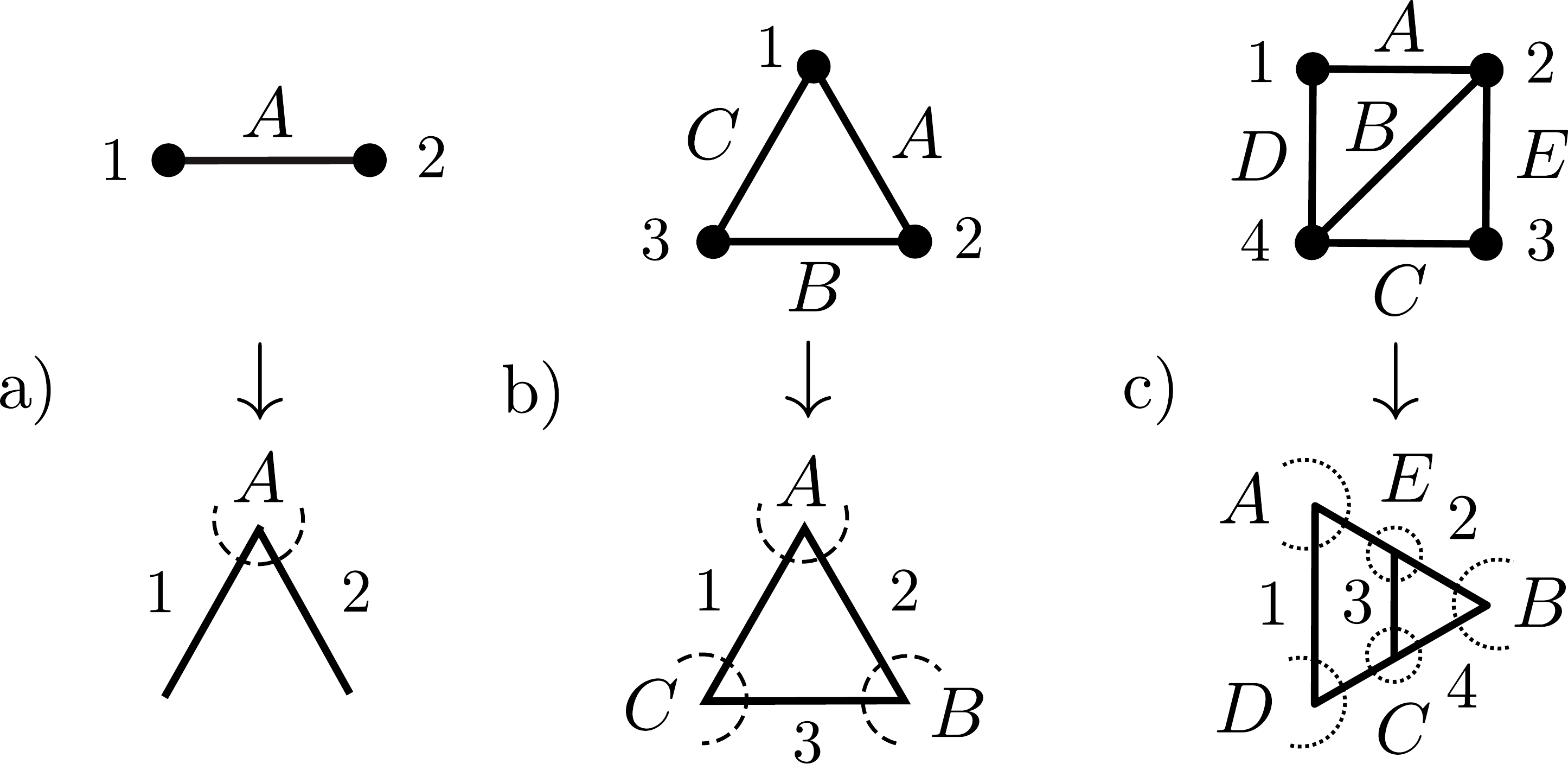}
\caption{Examples for the index conventions for Mayer and intersection
diagrams. For pairwise intersecting particles, the two representations are dual
to each other with the role of nodes and edges interchanged. Upper case letters
indicate Mayer bonds or intersection centers. Whereas particle numbers represent
root points or particle lines.}
\label{fig:1-2-3-coordinates}
\end{figure}

The simplest diagram, apart from the second virial cluster, is the triangle
graph, which has been discussed by Wertheim \cite{wertheim-1, wertheim-2,
wertheim-3, wertheim-4} and in \cite{korden-2}. Using the indices of
Fig.~\ref{fig:1-2-3-coordinates}b), its corresponding integral in weight
functions
\begin{align}\label{3-exact}
\beta_2 &=\frac{1}{2V} \int \; 
f_{12}f_{23}f_{31}\,\rho_1\rho_2\rho_3
\;d\gamma_1 d\gamma_2 d\gamma_3\nonumber \\
&= \frac{1}{2V} \int \;
C^{A_1A_2}\omega^1_{A_1}(\vec{r}_a-\vec{r}_1)
\omega^2_{A_2}(\vec{r}_a-\vec{r}_2) \nonumber\\[0.4em]
&\qquad\quad\times
C^{B_2B_3}\omega^2_{B_2}(\vec{r}_b-\vec{r}_2)
\omega^3_{B_3}(\vec{r}_b-\vec{r}_3)\\[0.4em]
&\qquad\quad \times \;C^{C_3C_1}\omega^3_{C_3}(\vec{r}_c-\vec{r}_3)
\omega^1_{C_1}(\vec{r}_c-\vec{r}_1)\nonumber\\[0.4em]
&\quad \times d^3r_a d^3r_b d^3r_c\,
\rho_1(\gamma_1)\rho_2(\gamma_2)\rho_3(\gamma_3)\;\,d\gamma_1 d\gamma_2
d\gamma_3\nonumber
\end{align}
is a functional of the particle densities. The same integral in 2-point
densities (\ref{k-point}) allows the more compact notation
\begin{align}\label{3-exact-2}
&\beta_2 =\frac{1}{2V} C^{A_1A_2} C^{B_2B_3} C^{C_3C_1}\\
&\times \int n_{A_1C_1}(\vec{r}_{ac}) n_{A_2B_2}(\vec{r}_{ab})
n_{B_3C_3}(\vec{r}_{bc})\, d^3r_{ab} d^3r_{bc} d^3r_{ca}\nonumber
\end{align}
using the distance vectors $\vec r_{ab} = \vec r_b - \vec r_a$.

In the following discussion neither the dependence on the particle densities nor
the loop constraints will be of relevance, so that the identity
(\ref{f-function}) can be written in the simplified form
\begin{equation}\label{short-1}
f_{12}(A) \widehat{=}  C^{A_1 A_2}\omega_{A_1}\omega_{A_2}\;,
\end{equation}
using the index combination
\begin{equation}\label{index-conv-comb}
\begin{split}
A_i\; : \quad & \text{weight index $A$ at intersection}\\
& \text{domain $\vec r_a$ of particle $i$}
\end{split}
\end{equation}
and omitting any reference to the integration over the intersection coordinates.
The index $A$ has now two meanings. On the one hand it indicates the
intersection center $A$, on the other hand it also numbers the link of the Mayer
cluster $f_{ij}(A)$, as seen in Fig.~\ref{fig:1-2-3-coordinates}a).

The two representations (\ref{3-exact}) and (\ref{3-exact-2}) of the three 
particle Mayer cluster of Fig.~\ref{fig:1-2-3-coordinates}b) can now be written
in the more convenient form:
\begin{align}\label{short-2}
& f_{12}(A)f_{23}(B)f_{31}(C)\\[0.4em]
& \widehat{=} C^{A_1A_2}\omega_{A_1}\omega_{A_2}\cdot
C^{B_2B_3}\omega_{B_2}\omega_{B_3}
\cdot C^{C_3C_1}\omega_{C_3}\omega_{C_1}\nonumber\\[0.4em]
&= C^{A_1A_2}C^{B_2B_3}C^{C_3C_1}(\omega_{A_1}\omega_{C_1})
(\omega_{A_2}\omega_{B_2})(\omega_{B_3}\omega_{C_3})\nonumber
\end{align}
with the definition of the 2-point function (\ref{k-point-func}) indicated by
parenthesis. Correspondingly, example Fig.~\ref{fig:1-2-3-coordinates}c) is the
product of 3-point and 2-point functions
\begin{equation}\label{short-3}
\begin{split}
&f_{12}(A)f_{23}(E)f_{34}(C)f_{41}(D)f_{24}(B)\\[0.4em]
&\widehat{=} C^{A_1A_2}C^{E_2E_3}C^{C_3C_4}C^{D_1D_4}C^{B_2B_4}\\[0.4em]
&\times (\omega_{A_1}\omega_{D_1})(\omega_{A_2}\omega_{B_2}\omega_{E_2})
(\omega_{C_3}\omega_{E_3})(\omega_{C_4}\omega_{B_4}\omega_{D_4})\;.
\end{split}
\end{equation}

\begin{figure}
\includegraphics[width=1.9cm,angle=-90]{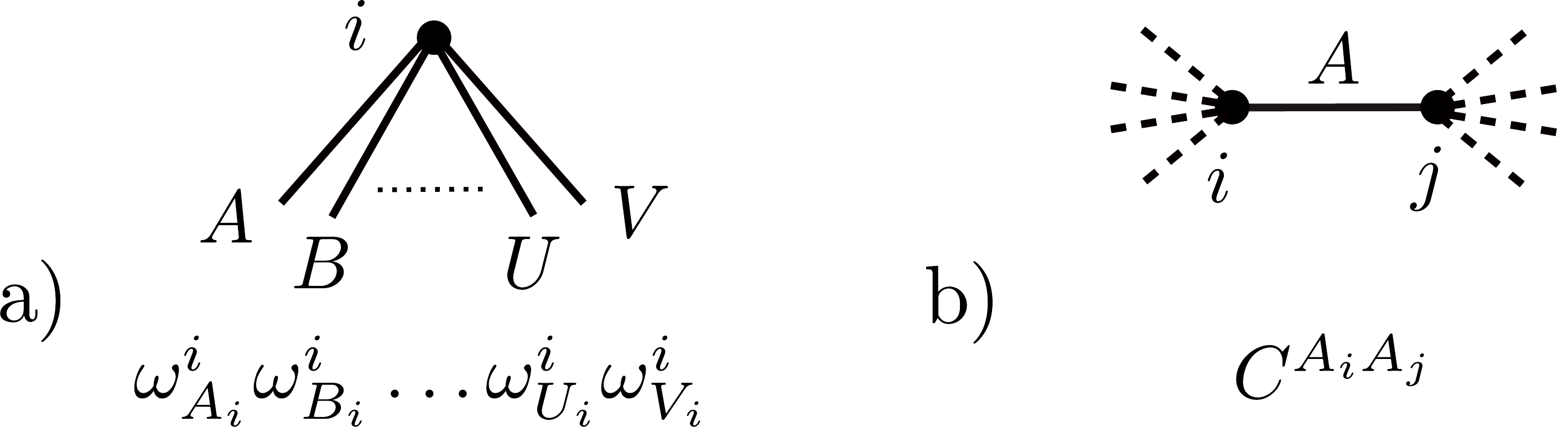}
\caption{Substitution rules for Mayer diagrams: a) The Mayer vertex of particle
$i$ with $k$ attached bonds corresponds to a $k$-point function and b) the Mayer
bond linking particles $i$ and $j$ maps to a $C$-matrix.}
\label{fig:Mayer-vertex}
\end{figure}

Comparing the three examples (\ref{short-1}), (\ref{short-2}), and
(\ref{short-3}) to their corresponding Mayer diagrams in
Fig.~\ref{fig:1-2-3-coordinates} reveals a simple building rule for virial
integrals represented in weight functions. Define the ''Mayer vertex`` as the
node of a Mayer diagram with its attached edges, as shown in
Fig.~\ref{fig:Mayer-vertex}a).
The cluster integral can then be read off directly from the Mayer graph by the
substitution
\begin{equation}\label{substitute}
\begin{split}
\text{Mayer vertex} & \to \text{$k$-point function}\\
\text{Mayer bond}  & \to \text{$C$-matrix}
\end{split}
\end{equation}
multiplied by the particle densities, loop constraints, integral measures, and
the symmetry factor. For the generic case of pairwise intersecting particles,
this simple replacement defines a unique mapping between the representation of
Mayer and intersection diagrams.

Graphically, the building rule (\ref{substitute}) corresponds to an exchange of
nodes and edges, as shown in Fig.~\ref{fig:1-2-3-coordinates}, which in terms of
graph theory defines the ''dual graph`` \cite{diestel} of the Mayer diagram.
However, this bijection still provides no simplification of the integral, and
the evaluation of the dual diagram is as complicated and unmanageable as it is
for the original virial cluster. The next step considers therefore the
systematic approximation of intersection diagrams of pairwise intersecting
particles. 
\subsection{Contraction rules}\label{subsubsec:contraction_rules}
It has already been shown in \cite{korden-2} that the structure of intersection
diagrams can be simplified by moving some of their intersection centers into a
single one. This process of ''contraction`` increases the rotational and
translational degrees of freedom over which the statistical system is averaged.
It therefore reduces the complexity of the virial integrals but also coarsen
their spatial resolution. For the most simplest cases, as the clusters of
Fig.~\ref{fig:four-virial} and Fig.~\ref{fig:zero-loop}, the contraction of
diagrams can be done by hand.

\begin{figure}
\includegraphics[width=3.2cm,angle=-90]{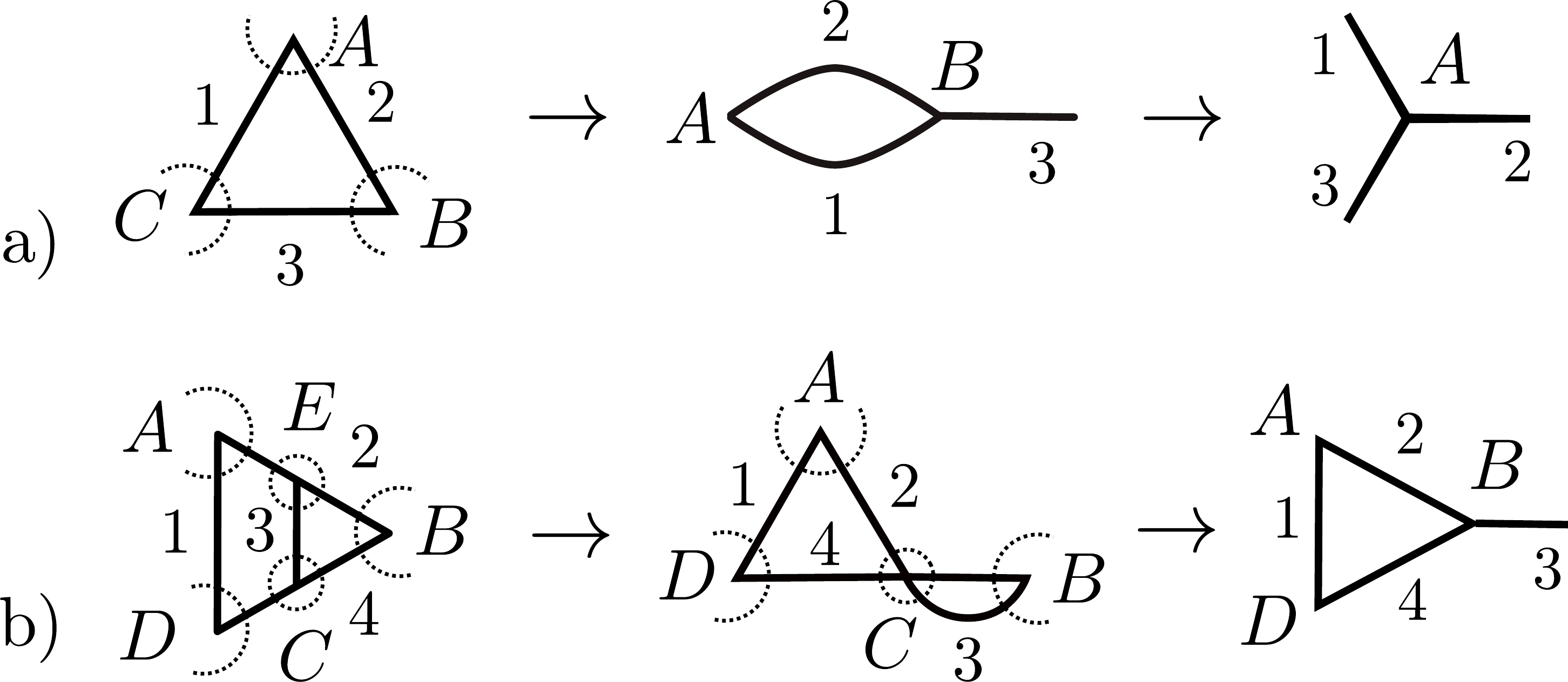}
\caption{The approximation scheme of FMT is based on the successive contraction
of intersection points. Reducible diagrams are intermediate steps with two
particles intersecting more than once. Further contractions finally result in
irreducible graphs.}
\label{fig:contraction}
\end{figure}

To derive an algebraic set of contraction relations, consider the pairwise
contraction of the triangle diagram shown in Fig.~\ref{fig:contraction}a).
The three intersection centers are combined in two steps: First, $C$ is shifted
into $B$, leaving the point $A$ invariant; then $B$ is shifted into $A$. For
each of the three diagrams of Fig.~\ref{fig:contraction}a) we can now write
down their corresponding virial integrand, using the splitting rules
(\ref{splitting_1}) and (\ref{splitting_2}) of the Euler form:
\begin{align}
&C^{A_1A_2}\, \omega_{A_1}\omega_{A_2}\cdot C^{B_2B_3}C^{C_3C_1}\,
\omega_{B_2}\omega_{B_3} \omega_{C_3}\omega_{C_1}\nonumber\\
&\xrightarrow{C \to B} C^{A_1A_2}\, \omega_{A_1}\omega_{A_2}\cdot 
C^{B_2B_3B_1}\, \omega_{B_2}\omega_{B_3} \omega_{B_1}\label{contract_1}\\
&\xrightarrow{B\to A} C^{A_1A_2A_3}\, \omega_{A_1}\omega_{A_2}
\omega_{A_3}\;.\label{contract_2}
\end{align}

The first rule of the contraction operation can now be read off from
(\ref{contract_1}) for the fusion of two intersection centers
\begin{equation}\label{contraction-e}
\begin{split}
\lim_{C\to B}&  C^{B_2B_3}C^{C_3C_1}\,
\omega_{B_2}\omega_{B_3}\omega_{C_3}\omega_{C_1}\\
=\; &C^{B_2B_3B_1}\,\omega_{B_2}\omega_{B_3}\omega_{B_1}\;.
\end{split}
\end{equation}

More clearly, the effect of the limit $C\to B$ on the weight functions and
$C$-matrices can be grouped into two classes. If the objects belong to different
particles
\begin{equation}\label{con_split_1}
\begin{split}
&\lim_{C\to B}\omega_{C_3} =  \omega_{B_3}\;, \quad 
\lim_{C\to B}\omega_{B_2}\omega_{C_3} = \omega_{B_2}\omega_{B_3}\\
&\lim_{C\to B} C^{C_3C_1} = C^{B_3B_1}\\
\end{split}
\end{equation}
the indices of the intersection centers are simply renamed. The same operation
for identical particle indices
\begin{equation}\label{con_split_2}
\lim_{C\to B} C^{B_2B_3}C^{C_3C_1}\,
\omega_{B_3}\omega_{C_3} = C^{B_2B_3B_1}\omega_{B_3}
\end{equation}
yields a tensorial contraction of the $C$-matrices, aligned with the removal of
one weight function. This operation is only defined in the combination of
weight functions and $C$-matrices and cannot be split in the way of
(\ref{con_split_1}). 

Applying these rules to Eq.~(\ref{contract_1})
\begin{equation}
\begin{split}
\lim_{B\to A} & C^{A_1A_2}C^{B_2B_3B_1}\, \omega_{A_1}\omega_{A_2} 
\omega_{B_2}\omega_{B_3} \omega_{B_1}\\
 =\; & C^{A_1A_2A_3}\, \omega_{A_1}\omega_{A_2}
\omega_{A_3}
\end{split}
\end{equation}
reproduces the graphically obtained result of (\ref{contract_2}), provided
Eq.~(\ref{con_split_2}) is extended to two identical particle indices. This
observation is readily generalized to the pairwise contraction of $n-k$
coincident particle indices
\begin{align}\label{contraction}
\lim_{B\to A} & C^{A_1\ldots A_k A_{k+1}\ldots A_n}\; C^{B_{k+1}\ldots B_n
B_{n+1}\ldots B_m}\nonumber\\
&\times \omega_{A_{k+1}}\ldots\omega_{A_n}\;\omega_{B_{k+1}}\ldots
\omega_{B_n}\\[0.4em]
=\; &  C^{A_1\ldots A_{k+1}\ldots A_n\ldots A_m}
\omega_{A_{k+1}}\ldots\omega_{A_n}\;.\nonumber
\end{align}

Thus the successive contraction of intersection centers generates higher rank
$C$-matrices. However, from the splitting rules of the Euler form
(\ref{splitting_1}) and (\ref{splitting_2}) we know that the maximal rank of a
$C$-matrix for 3-dimensional particles is at most 3 and that all further indices
necessarily reduce to the index $v$ of the particle volume. It is therefore
natural to combine the two equations (\ref{splitting_1}), (\ref{splitting_2})
into one
\begin{equation}
K(\partial \St_k) = C^{(A_1\ldots A_k)}\omega_{A_1}\ldots \omega_{A_k}
\end{equation}
and to define the generalized $C$-matrix 
\begin{equation}\label{c-matrix}
C^{(A_{i_1}A_{i_2}A_{i_3}A_{i_4}\ldots A_{i_k})} := 
C^{(A_{i_1}A_{i_2}A_{i_3}}\,\delta_{v_{i_4}}^{A_{i_4}}\ldots
\delta_{v_{i_k}}^{A_{i_k})}\;,
\end{equation}
where the parenthesis indicate the symmetrization of all particle indices. 

As an example, let us expand Eq.~(\ref{contract_2}) in the neutral element
$\omega_v$ for three identical particles:
\begin{align}
& C^{(A_1A_2A_3)} \, \omega_{A_1}\omega_{A_2} \omega_{A_3}\\
& = C^{(\chi_{i_1}v_{i_2} v_{i_3})} \, \omega_{\chi_{i_1}}\omega_{v_{i_2}}
\omega_{v_{i_3}} 
+C^{(\alpha_{i_1}\alpha_{i_2}v_{i_3})} \,
\omega_{\alpha_{i_1}}\omega_{\alpha_{i_2}} \omega_{v_{i_3}}\nonumber\\
&\quad  + C^{(\alpha_{i_1}\alpha_{i_2} \alpha_{i_3})} \,
\omega_{\alpha_{i_1}}\omega_{\alpha_{i_2}}
\omega_{\alpha_{i_3}}\nonumber\\
& = 3\,\omega_\chi\omega_v^2
+6\,C^{\alpha_1\alpha_2}\,\omega_{\alpha_1}\omega_{\alpha_2}\omega_v
+6\,C^{\alpha_1\alpha_2\alpha_3} \, \omega_{\alpha_1} \omega_{\alpha_2}
\omega_{\alpha_3}\;. \nonumber
\end{align}
The result correctly reproduces the Euler form (\ref{euler-stack}) for $k=3$.

The successive application of pairwise contractions on dual Mayer clusters
generates a vast number of diagrams. However, some of them correspond to
networks with multiple intersections between particles, as shown in  
Fig.~\ref{fig:contraction}a). The first step $C\to B$ generates an intermediate
diagram with the particles $1$ and $2$ intersecting twice in the centers of $A$
and $B$. But as one intersection point already determines the position and
orientation of their particles uniquely, this diagram is no allowed
configuration. Whereas the next contraction, $B\to A$, resolves this ambiguity.

Intermediate diagrams are identified as products of $C$-matrices with more than
one common particle index. To distinguish these cases from admissible
intersection diagrams, we introduce the notation:
\begin{equation}
\begin{split}
&C^{A_1\ldots A_k A_{k+1}\ldots A_n}\; C^{B_{k+1}\ldots B_n
B_{n+1}\ldots B_m}\\
&n-k = 1\;:\qquad \text{irreducible intersection}\\
&n-k > 1\;:\qquad \text{reducible intersection.}
\end{split}
\end{equation}
Correspondingly, diagrams without reducible intersections are referred to as
''irreducible diagrams`` and ''reducible diagrams`` otherwise. It follows from
their definition that any reducible intersection can be reduced to an
irreducible one by further contractions. 

Another example is the Mayer diagram of Fig.~\ref{fig:1-2-3-coordinates}c). Its
contractions can be either derived by (\ref{contraction}) or read off from
Fig.~\ref{fig:contraction}b)
\begin{align}
&C^{A_1A_2}C^{B_2B_4}C^{D_1D_4}C^{C_3C_4}C^{E_2E_3}\nonumber\\
&\qquad \times \omega_{A_1}\omega_{A_2}
\omega_{B_2}\omega_{B_4}\omega_{D_1}\omega_{D_4}
\omega_{C_3}\omega_{C_4}\omega_{E_2}\omega_{E_3} \nonumber\\
&\xrightarrow{E \to
C}C^{A_1A_2}C^{B_2B_4}C^{D_1D_4}C^{C_2C_3C_4}\label{contract_3}\\
&\qquad \times \omega_{A_1}\omega_{A_2}
\omega_{B_2}\omega_{B_4}\omega_{D_1}\omega_{D_4}
\omega_{C_2}\omega_{C_3}\omega_{C_4}\nonumber\\
&\xrightarrow{C\to B}C^{A_1A_2}C^{D_1D_4}C^{B_2B_3B_4}\label{contract_4}\\
&\qquad \times \omega_{A_1}\omega_{A_2}
\omega_{D_1}\omega_{D_4}\omega_{B_2}\omega_{B_3} \omega_{B_4}\;.\nonumber
\end{align}
The first contraction, $E\to C$, yields again an intermediate diagram
(\ref{contract_3}), reducible in the particle numbers $2$ and $4$, which is
then transformed into an irreducible one by shifting $C\to B$. The result is
the highest possible approximation of the Mayer cluster of
Fig.~\ref{fig:1-2-3-coordinates}c). No further contraction is possible as the
particles $1$ and $3$ do not interact directly. This can be seen either from
the Mayer diagram, where the corresponding $f_{13}$ is missing, or directly
from the intersection diagram of Fig.~\ref{fig:contraction}b). Therefore,
apart from irreducibility, the intersection diagrams obtained by pairwise
contractions also have to be compatible to the bonding relations of its Mayer
graph. 

Again, compatibility of a diagram can be directly read off from its
corresponding Mayer cluster, as each of the generalized $C$-matrices belongs to
a completely connected Mayer subdiagram:
\begin{equation}\label{compatibility}
\begin{split}
C^{(A_1\ldots A_k)}\;\widehat{=}\;&\text{completely connected Mayer}\\
& \text{subdiagram of the particles $1, \ldots, k$.}
\end{split}
\end{equation}
The diagram Fig.~\ref{fig:1-2-3-coordinates}c) from the previous example can
therefore be contracted either to $C^{B_2B_3B_4}$ or $C^{B_1B_2B_4}$,
corresponding to their subdiagrams of particle indices $(2,3,4)$ and $(1,2,4)$.

In the next subsection it will be shown that any diagram can be decomposed into
its maximally connected subgraphs. In most cases this splitting is uniquely
defined. But the current example is one of the exceptional cases, which can be
split in at least two different ways, corresponding to a global $\mathbb{Z}_2$
symmetry. For such diagrams, it is necessary to count their multiplicity of
contractions, indicated by 
\begin{equation}\label{m}
 m : \text{contraction multiplicity.}
\end{equation}
For the graph of Fig.~\ref{fig:1-2-3-coordinates}c), the multiplicity is
therefore 
\begin{equation}\label{contr-multi}
m(1,0,0)=2\quad \text{and}\quad m(n_1,n_2,n_3)=1\quad \text{else.} 
\end{equation}
\subsection{Contraction rules and their algebra}
\label{subsubsec:algebra}
The contraction rules are local mappings on the set of Mayer and intersection
diagrams. In order to describe their operations on general graphs, it is
practical to introduce a suitable notation for both types of representations.

As only completely connected subdiagrams can be contracted into single
intersection centers, they take up the position of prime elements in the set of
Mayer clusters and intersection diagrams. Let us therefore introduce the
notation:
\begin{align}\label{subdiagram_M}
\Gamma_n^\lambda\;:&\; \text{completely connected 
Mayer subdiagram of $n$}\nonumber \\
&\; \text{particles and external bonds grouped in the}\nonumber\\
&\; \text{partition $\lambda$.}\nonumber
\end{align}
Due to the permutation symmetry of the particle indices of completely connected
diagrams, it is sufficient to group the external links into a partition table
$\lambda$. For example, $\lambda=[(ABC),(DEFG)]$ assigns the external links
$ABC$ to subdiagram $1$ and $DEFG$ to subdiagram $2$.

Any Mayer cluster can now be represented as a product of prime subdiagrams. For
example, Fig.~\ref{fig:1-2-3-coordinates}c) allows the two decompositions:
\begin{equation}
\Gamma_3^{(ABD)CE}\,\Gamma_1^{CE} = \Gamma_3^{(BCE)AD}\,\Gamma_1^{AD}\;,
\end{equation}
which directly translates to the splitting of the Euler form
\begin{equation}
K(\Gamma_3^{(ABD)CE}\,\Gamma_1^{CE}) = K(\Gamma_3^{(ABD)CE})\,K(\Gamma_1^{CE})
\end{equation}
This notation is far more compact than the representation in weight functions
(\ref{contract_3}), (\ref{contract_4}). 

Analogously, intersection diagrams can be split at each intersection center
along their particle lines:
\begin{equation}\label{subdiagram_I}
\begin{split}
\widetilde \Gamma_A^\lambda\;:&\;
\text{intersection subdiagram of center $A$ and}\\
&\; \text{partition $\lambda$ of external particle lines.}
\end{split}
\end{equation}
As an example, the intersection diagram of Fig.~\ref{fig:1-2-3-coordinates}c)
factorizes into the prime elements
\begin{equation}
\widetilde\Gamma_A^{12}\, \widetilde\Gamma_E^{23}\, \widetilde\Gamma_B^{24}
\widetilde\Gamma_C^{34}\, \widetilde\Gamma_D^{41}\;.
\end{equation}

In summary, the derivation and approximation of intersection diagrams reduces
to a simple set of operations and constraints. In combination with the splitting
rules of the Euler form and its linearity, we now define the ''intersection
algebra``:
\begin{definition}\label{def:inter_algebra}
Let $K$ be the Euler form and $\Gamma_{n,b}\in \Gamma$ an element of the set
of Mayer star-clusters $\Gamma$ with $n$ particles, factorizing into the prime
subdiagrams $\Gamma_m^\lambda$
\begin{equation}
\Gamma_n = \prod_i \Gamma_{n_i}^{\lambda_i}\;, \quad n=\sum_i n_i\;.
\end{equation}
The Euler form induces a real, linear operation on the set of Mayer clusters
\begin{equation}\label{euler-mapping}
\begin{split}
K(x_1\Gamma_{n_1} + x_2\Gamma_{n_2}) &= x_1 K(\Gamma_{n_1}) +
x_2 K(\Gamma_{n_2})\\
K(\Gamma_n) &= \prod_i K(\Gamma_{n_i}^{\lambda_i})
\end{split}
\end{equation}
for $x_1,x_2\in \mathbb{R}$, defining the ''intersection algebra`` $(K,
\Gamma)$. The splitting relation
\begin{equation}\label{rep-euler}
K(\Gamma_n^A) = C^{(A_1A_2\ldots A_n)}\,\omega_{A_1}\omega_{A_2}\ldots 
\omega_{A_n}
\end{equation}
induces a representation on $\Gamma$ in weight functions. Intersection centers
can be combined by pairwise contractions 
\begin{equation}
C : \lambda \to \lambda'\;,\quad |\lambda| > |\lambda'|\;,
\end{equation}
reducing the length of the partition $|\lambda|$ by at least one.
\end{definition}

The intersection algebra changes the focus from differential geometry to
the representation theory of the symmetric group \cite{reps, sachkov}, with the
Euler form (\ref{rep-euler}) relating the partition table $\lambda$ of an
intersection diagram to the ring of symmetric polynomials. To prove that the
contraction operation respects this representation, apply the relation
\begin{equation}
(\sum_{i=0}^\infty a_i x^i)(\sum_{j=0}^\infty b_j x^j)= \sum_{n=0}^\infty
(\sum_{k=0}^n a_k b_{k-n})x^n
\end{equation}
to the product of two generating functions, contracted in the first particle
index $i_1$:
\begin{align}\label{ring}
\lim_{B\to A} & (\sum_n C^{(A_1A_2\ldots A_n)} \omega_{A_1}^{i_1}
\omega_{A_2}^{i_2} \ldots \omega_{A_n}^{i_n} \rho_{i_2} \ldots \rho_{i_n})
\nonumber\\
&\times (\sum_k C^{(B_1B_2\ldots B_k)} \omega_{B_1}^{i_1} \omega_{B_2}^{j_2}
\ldots \omega_{B_k}^{j_k} \rho_{j_2}\ldots \rho_{j_k})\;\rho_{i_1}\nonumber\\
&= \sum_n n\, C^{(A_1\ldots A_n)} \omega_{A_1}^{i_1} \ldots \omega_{A_n}^{i_n}
\rho_{i_1} \ldots \rho_{i_n}\;,
\end{align}
which again corresponds to the Euler form of a new intersection diagram. 

The ring structure of the polynomial (\ref{rep-euler}) greatly simplifies the
following derivation of the symmetry factors as well as the construction of
vertex functions. 
\section{Intersection vertices and vertex functions}\label{subsec:vertex}
The mapping (\ref{euler-mapping}) uniquely defines the splitting of any
intersection diagram into its Euler forms. In principal, this is all one needs
to derive higher order corrections of the free energy. However, much of this
approach can be simplified by the resummation of subdiagrams. In the following
two subsections, the free-energy representation in intersection centers will be
generalized. First, it will show in \ref{subsubsec:splitting} that the
functional splits locally at each intersection center into vertex functions,
whose analytical form will be derived in \ref{subsubsec:vertex}
\subsection{The local splitting of the FMT functionals}
\label{subsubsec:splitting}
In the notation of the completely connected intersection diagrams
(\ref{subdiagram_I}), the Rosenfeld functional is representable as the infinite
sum
\begin{equation}\label{euler-dig}
\Phi_{0,1}(\vec r_A) = \int \sum_{n=1}^\infty \frac{1}{n} 
K(\widetilde \Gamma_A^{i_1,\ldots i_n})\,\rho_{i_1}\ldots \rho_{i_n}\,dn_v
\end{equation}
over the subclass of diagrams with only one intersection center. An even more
concise notation can be obtained using the linearity of the Euler form
(\ref{euler-mapping}) and considering the weighted sum over diagrams 
\begin{equation}\label{ros-sum-1}
\sum_{n=1}^\infty \frac{1}{n}\, \widetilde \Gamma_A^{(i_1\ldots i_n)}\;,
\end{equation}
symmetrizised over the external particle indices. This shortened notation
provides a convenient representation for the discussion of diagrammatic
resummation. 

But the Euler form is only one aspect in the derivation of the free energy. In
the following we will also need to generalize the expansion of the functional in
intersection centers (\ref{func-def}) and to determine their combinatorial
prefactors (\ref{virial}) of the virial contributions.

Let us first focus on the expansion of the functional itself. Explicitly
written out up to three intersection centers
\begin{align}\label{free-expansion}
& \beta F =\int \left.\frac{\delta (\beta F)}{\delta n_v(\vec r_a)}
\right|_{g=0} \, \delta n_v(\vec r_a)\nonumber\\
& + \left.\frac{\delta^3 (\beta F)}{\delta n_v(\vec r_a) \delta n_v(\vec r_b)
\delta n_v(\vec r_c)}\right|_{g=1} \delta n_v(\vec r_a) \delta n_v(\vec r_b)
\delta n_v(\vec r_c)\nonumber\\[0.75em]
&+ \ldots \nonumber\\
&= \int \Phi_{0,1}\;d^3r_a + \Phi_{1,3}\;d^3r_a d^3r_b d^3r_c + \ldots\;,
\end{align}
it reproduces the integral representation of the  Rosenfeld functional
(\ref{ros-resum}) at first order and yields the next to leading order
\begin{equation}\label{leading_order}
\begin{split}
\Phi_{1,3} = \int \sum_{k=3}^\infty \frac{1}{k}&  \beta_{k-1}^{(1,3)}(\vec r_a,
\vec r_b, \vec r_c)\\
&\times  dn_v(\vec r_a)dn_v(\vec r_b)dn_v(\vec r_c)\;,
\end{split}
\end{equation}
which parallels the structure of (\ref{3-exact}) for $k=3$.

From the local property of the Euler form (\ref{euler-mapping}) follows that the
virial contribution factorizes into a product of three polynomials, each
depending on one of $n_v(\vec r_a), n_v(\vec r_b), n_v(\vec r_c)$. Consequently,
the integration of (\ref{leading_order}) factorizes likewise and can be executed
for each intersection center individually. The same argument applies of course
to all further terms of the expansion (\ref{free-expansion}). The free energy
functional decouples into a product of local functionals for each intersection
center, coupled only by the loop constraints and the numerical prefactors of the
virial integrals.

Given the irregular structure of the symmetry factors $\sigma(\Gamma)$ entering
(\ref{virial-integral}) for different diagrams, it is nontrivial that such a
splitting of the functionals $\Phi_{g,h}$ should exists. A general proof of this
hypothesis would require a classification of the automorphism groups of Mayer
clusters under relabeling, which to the best of our knowledge is unknown. We
will therefore focus on the leading correction term of the Rosenfeld functional
and explicitly derive (\ref{leading_order}) in the following sections. 

The diagrams entering $\Phi_{1,3}$ have three intersection centers grouped into
\begin{figure}
\includegraphics[width=4.5cm,angle=-90]{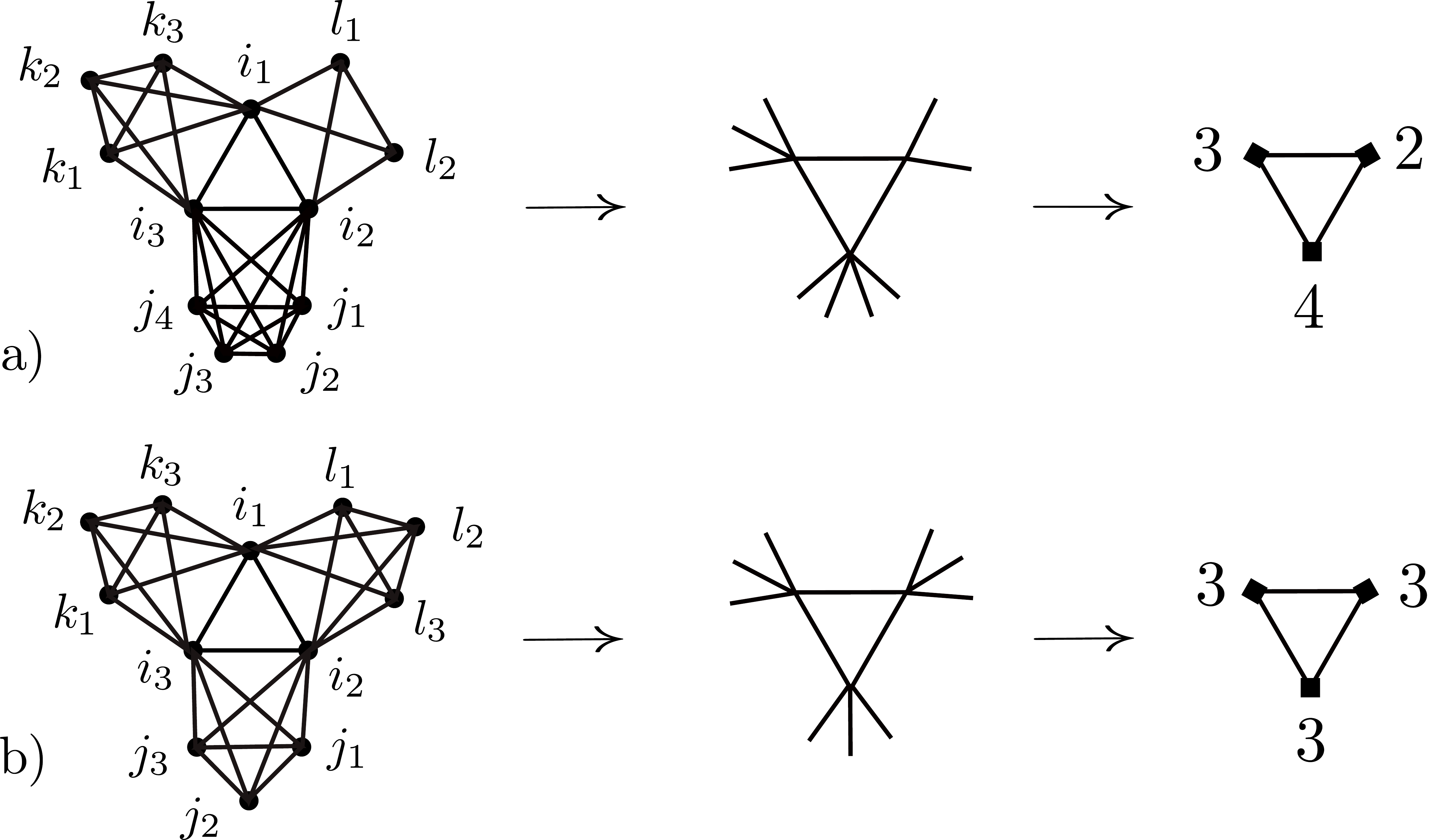}
\caption{Two examples of the class of triangular diagrams, represented as
Mayer graph, maximally contracted intersection diagram, and as weighted graph.
Their automorphism groups are: a) $\widetilde\Gamma(4,3,2)$: $E_1^3\times
S_4\times S_3\times S_2$ and b) $\widetilde\Gamma(3,3,3)$: $G_{12}$.}
\label{fig:butterfly}
\end{figure}
a ring and thus have a triangular substructure. Two examples are shown in
Fig.~\ref{fig:butterfly}, using the three representations as a Mayer graph,
as the maximally contracted intersection diagram, and as a weighted diagram with
the number of external lines as index. Using the notation of intersection
subdiagrams (\ref{subdiagram_I}), any such element can be written as
\begin{equation}\label{gamma_3}
\begin{split}
\widetilde \Gamma(n_1,n_2,n_3) &:= 
\widetilde \Gamma_{A}^{i_1i_2j_1\ldots j_{n_1}}
\widetilde \Gamma_{B}^{i_2i_3k_1\ldots k_{n_2}}
\widetilde \Gamma_{C}^{i_3i_1l_1\ldots l_{n_3}}\\
&= (\widetilde \Gamma_{A}^{n_1})^{i_1i_2}
(\widetilde \Gamma_{B}^{n_2})^{i_2i_3}
(\widetilde \Gamma_{C}^{n_3})^{i_3i_1}\;,
\end{split}
\end{equation}
with the paired indices $i_1,i_2,i_3$ defining the backbone of the graph and
$j,k,l$ indicating its external lines.

As shown in appendix \ref{sub:appendix_a}, the symmetry factor of a Mayer
diagram $\Gamma_n$ of $n$ particles is determined by the quotient
\begin{equation}\label{sym_mayer}
\sigma(\Gamma_n) = \frac{|S_n|}{|\text{Aut}(\Gamma_n)|}
\end{equation}
of the dimensions of the symmetric group $S_n$ and the automorphism group
$\text{Aut}(\Gamma_n)$ of the graph \cite{uhlenbeck-ford-1}. This remains true
for the dual diagram $\widetilde\Gamma$ as the invariance group is independent
of the representation of $\Gamma$ and therefore unaffected by contractions
\begin{equation}
\text{Aut}(\Gamma_n) = \text{Aut}(\widetilde \Gamma_n)\;.
\end{equation}
However, our definition of the particle stack (\ref{stack}) and its
representation as symmetrized $C$-matrix (\ref{c-matrix}) already includes
the invariance group of the completely connected subdiagrams. It is therefore
necessary to define an ''effective`` symmetry factor for intersection diagrams
$\widetilde\sigma( \widetilde\Gamma_n)$ without the invariance group of external
particle-lines. 

As an example, consider the graph $\widetilde\Gamma(3,3,3)$, shown in 
Fig.~\ref{fig:butterfly}b). Each of the three intersection centers has $5$
particle-lines attached, of which $2$ are fixed by the triangular backbone
diagram, whereas the remaining $3$ external lines are invariant under the
permutation $S_3$. In order to compensate for the symmetry of the $C$-matrices,
the group $S_3\times S_3\times S_3=(S_3)^3$ has to be factored out from
$\text{Aut}(\widetilde\Gamma(3,3,3))=G_{12}$:
\begin{equation}
G_{12}/(S_3)^3 = D_6\;.
\end{equation}
The quotient group is therefore the automorphism group of the weighted graph
shown in Fig.~\ref{fig:butterfly}b). As demonstrated in appendix
\ref{sub:appendix_a}, this result applies to any diagram
$\widetilde\Gamma(n,n,n)$, independent of the number of external
particle-lines. 

Generally, it is far easier to derive the reduces invariance group from the
weighted diagrams, where the external particle-lines of the intersection graph
have been replaced by their number as weight index. Using the results of 
Tab.~\ref{tab:group_all}, the ''reduced automorphism groups`` of triangular
diagrams $\widetilde \Gamma(n_1,n_2,n_3)$ consists of only four cases:
\begin{equation}\label{reduced_aut_group}
\begin{split}
n_1\neq n_2 \neq n_3 & :\qquad E_1^3 \\
n_1 = n_2 \neq n_3   &: \qquad E_1 \times \mathbb{Z}_2 \\
n_1 =  n_2 = n_3     &: \qquad D_6 \\
n_1=n_2 = 0, n_3=1   &: \qquad S_2\times\mathbb{Z}_2\;, 
\end{split}
\end{equation}
with the exceptional diagram $\widetilde\Gamma(1,0,0)$ first discussed in the
context of the contraction rules in Sec.~\ref{subsubsec:contraction_rules}. 

The reduced invariance group identifies four equivalence classes of diagrams,
independent of their particle numbers. This is an important property, as the
final goal of determining the free energy functional $\Phi_{1,3}$ requires the
resummation of all such diagrams. And here we see that this problem reduces to
at most four different classes.

Instead of the symmetry factor for virial integrals (\ref{sym_mayer}),
$\widetilde\sigma$ depends only on the equivalence classes of diagrams and is
independent of their particle numbers. In the representation of intersection
diagrams (\ref{subdiagram_I}), the number of external, unpaired particle-lines
can be determined from the partition table $\Lambda$:
\begin{equation}
\widetilde \Gamma_n^\Lambda = \prod_{i=1}^k \widetilde
\Gamma_{A_i}^{\lambda_i}\quad ,\quad 
\Lambda = \{\lambda_1,\ldots \lambda_k \}\;.
\end{equation}
Let $(\lambda_i)_j \in \lambda_i$ denote an individual element of the
partition $\lambda_i \in \Lambda$. The set of external lines is then
characterized as:
\begin{equation}
\lambda^\perp_i := \{\;(\lambda_i)_j\;|\; (\lambda_i)_j\in \lambda_i\;,\;  
(\lambda_i)_j\not\in \lambda_k\;,\; \forall\; i\neq k\}\;.
\end{equation}
Applied to the triangular diagrams (\ref{gamma_3}), the partition table
$\Lambda=(\lambda_A,
\lambda_B, \lambda_C)$ consists of the three elements
\begin{align}
\lambda_A &= ((i_1, i_2), (j_1,\ldots , j_{n_1-2}))\;,\nonumber\\
\lambda_B &= ((i_2, i_3), (k_1,\ldots, k_{n_2-2}))\;,\\
\lambda_C &= ((i_3, i_1),(l_1,\ldots, l_{n_3-2}))\;,\nonumber
\end{align}
from which follows the corresponding orthonormal set of unpaired indices
\begin{equation}
\begin{split}
\lambda_A^\perp = (j_1,\ldots &, j_{n_1-2})\;,\;
\lambda_B^\perp = (k_1,\ldots, k_{n_2-2})\;,\\
\lambda_C^\perp & = (l_1,\ldots, l_{n_3-2})\;.
\end{split}
\end{equation}

Using this notation, we define the effective symmetry factor of intersection
diagrams:
\begin{equation}\label{sym-i}
\begin{split}
\widetilde\sigma(\widetilde \Gamma_n) 
& := \frac{1}{n!}\,\sigma(\Gamma_n)\,\prod_i |\lambda_i^\perp|\,!\\
&\;=\frac{1}{|\text{Aut}(\Gamma_n)|}\,\prod_i
|\lambda_i^\perp|\,!\;, 
\end{split}
\end{equation}
which determines the inverse of the dimension of the reduced invariance group
and therefore depends only on their equivalence classes of diagrams. Its values
for the triangular diagrams are listed in Tab.~\ref{tab:group_all}.

This definition of the symmetry factor not only compensates for the symmetry
of the $C$-matrix, but also includes the normalization factor $1/(n-1)!$ of the
virial coefficient (\ref{virial-integral}) and the $1/n$ of its integral
(\ref{virial_old}). All numerical prefactors of the free-energy virial expansion
are thus included in $\widetilde\sigma$.
\subsection{Vertex functions}\label{subsubsec:vertex}
The new symmetry factor $\widetilde\sigma$ does not affect the previous
derivation of the Rosenfeld functional (\ref{beta-1}), (\ref{ros-2}). Its
numerical value for the starfish diagrams with unpaired particle-lines
$\lambda_i = \lambda_i^{\perp}$
\begin{equation}
\widetilde\sigma(\widetilde\Gamma_A^{i_1\ldots i_n})
= \frac{1}{|S_n|}\;n! = 1
\end{equation}
coincides with the corresponding symmetry factor of the Mayer diagram
$\sigma(\Gamma_n^A)=1$. The structure of the resummation of 0-loop diagrams
(\ref{ros-sum-1}) remains therefore the same:
\begin{equation}\label{resum-vertex-1}
\begin{split}
\sum_{n=1}^\infty \frac{\sigma(n)}{n!}\widetilde\Gamma_A^{i_1\ldots 
i_n} & = \sum_{n=1}^\infty \frac{\widetilde\sigma(n)}{n}
\widetilde\Gamma_A^{(i_1\ldots i_n)}\\
& = \sum_{n=1}^\infty \frac{1}{n}\widetilde\Gamma_A^{(i_1\ldots i_n)}\;.
\end{split}
\end{equation}

\begin{figure}
\includegraphics[width=2.0cm,angle=-90]{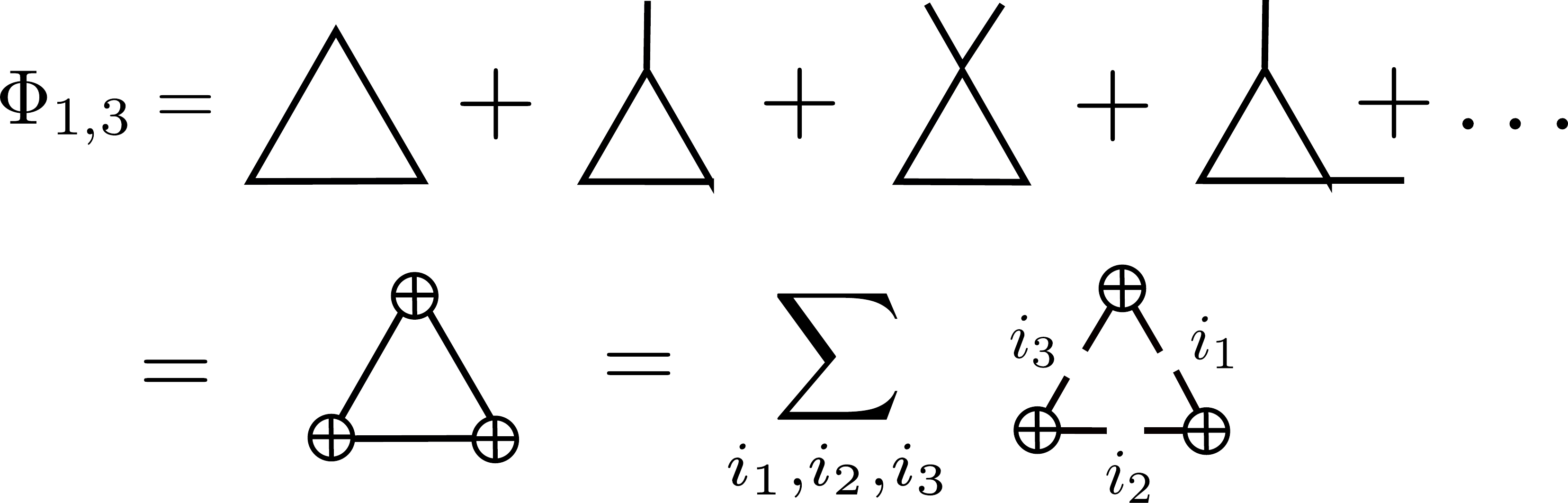}
\caption{Resummation of the triangular diagrams provides the first order
correction to the Rosenfeld functional. With the intersection centers indicated
by crossed circles, the functional is the product of three 2-vertex functions
summed over the inner particle indices $i_1,i_2,i_3$.}
\label{fig:resummation}
\end{figure}

To get a first impression, how this result generalizes to diagrams with more
than one intersection center, consider Fig.~\ref{fig:resummation}.
The first element of the series of triangular diagrams is the exact third virial
integral, which later will replace the approximate term in the Rosenfeld
functional. The next element is the contracted diagram of $\widetilde
\Gamma(1,0,0)$, shown in Fig.~\ref{fig:four-virial}b) and obtained by attaching
one additional external particle line to one of the three intersection centers
$A,B,C$. Iterating this operation on the backbone diagram, it generates the
series of $\Phi_{1,3}$, which can be resummed in the same way as the 0-loop
diagrams, shown in Fig.~\ref{fig:zero-loop}. For each intersection center we
thus obtain a ''vertex function`` $A\to V_{i_1i_2}(A)$ depending on the
particle-indices $i_1, i_2, i_3$. 

From an algebraic point of view, this factorization of the functional is a
consequence of the decomposition of the triangular diagram (\ref{gamma_3}) into
subdiagrams. However, although uniquely defined for a given set of particle
indices, it is invariant under permutation of the intersection centers $A,B,C$.
Instead of the single Mayer graph $\Gamma(1,0,0)$, there exists three identical
intersection diagrams:
\begin{equation}\label{diff-split}
\widetilde\Gamma(1,0,0)
= \widetilde\Gamma_A^2\widetilde\Gamma_B^1\widetilde\Gamma_C^1
= \widetilde\Gamma_A^1\widetilde\Gamma_B^2\widetilde\Gamma_C^1
= \widetilde\Gamma_A^1\widetilde\Gamma_B^1\widetilde\Gamma_C^2\;.
\end{equation}
In order to compensate for identical products of subdiagrams, let us introduce
the 
\begin{equation}
p : \text{polynomial multiplicity.}
\end{equation}

For the current product of three vertex functions, the combinatorial factor
$p(n_1,n_2,n_3)$ for $\widetilde \Gamma(n_1,n_2,n_3)$ derives from the binomial
coefficient of the generating function:
\[ 
\left( \sum_{n=1}^\infty x_n\right)^3 = 
\sum_{k_1,k_2\ldots=0}^3 \binom{3}{k_1,k_2,k_3,\ldots}x^{k_1}_1
x^{k_2}_2 x^{k_3}_3 \ldots
\]
reducing to the three representative cases of the index-vector $\vec k =
(k_1,k_2,\ldots)$:
\begin{equation}\label{poly_mult}
\begin{split}
(1,1,1) : \quad n_1\neq n_2 \neq n_3 & :\quad p=6 \\
(2,1,0) : \quad n_1 = n_2 \neq n_3   &: \quad p=3 \\
(3,0,0) : \quad n_1 = n_2 = n_3      &: \quad p=1
\end{split}
\end{equation}

Taking this into account and the contraction multiplicity $m$ defined in
(\ref{m}), the generating function $\Phi_{1,3}$ reduces to the product of
three vertex functions:
\begin{align}
&\Phi_{1,3}(\Gamma) =\sum_{k=0}^\infty\, \frac{\sigma(\Gamma_{k+3})}{(k+3)!}\,
\Gamma_{k+3}\label{phi-1}\\
&=\hspace{-1em}\sum_{n_1,n_2,n_3=1}^\infty
\sum_{i_1,i_2,i_3}\frac{\sigma(n_1,n_2,n_3)}{(n_1+n_2+n_3+3)!}\,
n_1!n_2!n_3!\label{phi-2} \\
&  \times \frac{m(n_1,n_2,n_3)}{p(n_1,n_2,n_3)}
(\frac{1}{n_1!}\widetilde\Gamma^{n_1}_{A})_{i_1i_2}
(\frac{1}{n_2!}\widetilde\Gamma^{n_2}_{B})_{i_2i_3}
(\frac{1}{n_3!}\widetilde\Gamma^{n_3}_{C})_{i_3i_1}\nonumber\\
& = \widetilde\kappa \sum_{i_1,i_2,i_3}
V(A)_{i_1i_2}V(B)_{i_2i_3}V(C)_{i_3i_1}\;,\label{phi-3}
\end{align}
with the constant $\widetilde\kappa$ to be derived in the following.

This result summarizes the central idea of the current work and requires some
commends. First, the structure of (\ref{phi-1}) is uniquely defined by the
virial expansion of the free energy (\ref{virial_old}) and the requirement that
its Euler form, suitably multiplied by powers of $\rho$ and integrated over the
particle coordinates, yields the free-energy functional density
(\ref{leading_order}). In the next line (\ref{phi-2}), the representation
(\ref{gamma_3}) of the triangular diagrams has been inserted, with only the
indices of the backbone diagram $i_1,i_2,i_3$ explicitely written out. After
exchanging the sums in (\ref{phi-3}), the vertex functions are a convenient
abbreviation for the resummed subdiagrams
\begin{equation}\label{vertex-m}
V_{i_1i_2}(A) := \sum_{n=1}^\infty
\frac{1}{n!}(\widetilde\Gamma_A^n)_{i_1i_2}\;.
\end{equation}

The definition of the vertex function agrees with (\ref{ros-sum-1}) and
introduces the factorials in (\ref{phi-2}) necessary to replace $\sigma$ by
$\widetilde \sigma$. The prefactor $\widetilde\kappa = \widetilde \sigma m/p$
is therefore a function on the equivalence classes of triangular diagrams and
can be evaluated by inserting (\ref{reduced_aut_group}), (\ref{poly_mult}), and
(\ref{contr-multi}). The detailed calculation can be found in appendix
\ref{sub:appendix_a} and yields the overall constant:
\begin{equation}\label{kappa}
\widetilde \kappa(n_1,n_2,n_3) = \frac{1}{6}\quad\text{for all}\quad 
\widetilde\Gamma(n_1,n_2,n_3)\;,
\end{equation}
independent of the particle numbers or equivalence classes. Instead, the
prefactor corresponds to its coefficient of the backbone diagram
$\widetilde\kappa(0,0,0)$.

The splitting of intersection diagrams into subdiagrams restricts the possible
dependence of $\widetilde\kappa$ on the number of external particle-lines.
Because of the invariance of the intersection centers under permutations, the
prefactor is either a function of the sum of particle numbers $f(n_1+n_2+n_3)$,
its product $g(n_1)g(n_2)g(n_3)$ or the product of both. For a general diagram,
whose backbone diagram of pairwise intersecting particles has $k$ intersection
centers $\widetilde\Gamma(n_1,\ldots, n_k)$, this generalizes to functions
invariant under the automorphism group
\begin{equation}
\widetilde\kappa(n_1,\ldots, n_k):\; g\widetilde\kappa = \widetilde\kappa
\;\;\text{for}\;\; g\in \text{Aut}(\widetilde\Gamma(0,\ldots,0))\;,
\end{equation}
with $\widetilde\Gamma(0,\ldots,0)$ as the intersection diagram with all
external particle-lines removed. This indicates, but not proves, that the
free-energy contribution for a given backbone diagram can be represented by
vertex functions that only depend on the number of paired particle-lines. The
prefactor for all diagrams coincide and correspond to $\widetilde
\Gamma(0,\ldots,0)$.

Translating back the diagrammatic representation (\ref{vertex-m}) of the vertex
function to the density dependent functional, we define the ''vertex function``
of $k$ internal particle lines
\begin{align}\label{vertex-i}
&V^{i_1\ldots i_n}(\vec r_a) \\
&= \int \sum_{m=0}^\infty \frac{1}{m+n} 
K(\widetilde \Gamma_A^{(i_1\ldots i_n k_1\ldots k_m)})\,
\rho_{k_1}\ldots \rho_{k_m}\, dn_v\nonumber
\end{align}
as the sum of Euler forms over prime subdiagrams. Here we also used the
observation of (\ref{free-expansion}) that the integration over $n_v$ factorizes
for each intersection center.
\begin{figure}
\includegraphics[width=1.5cm,angle=-90]{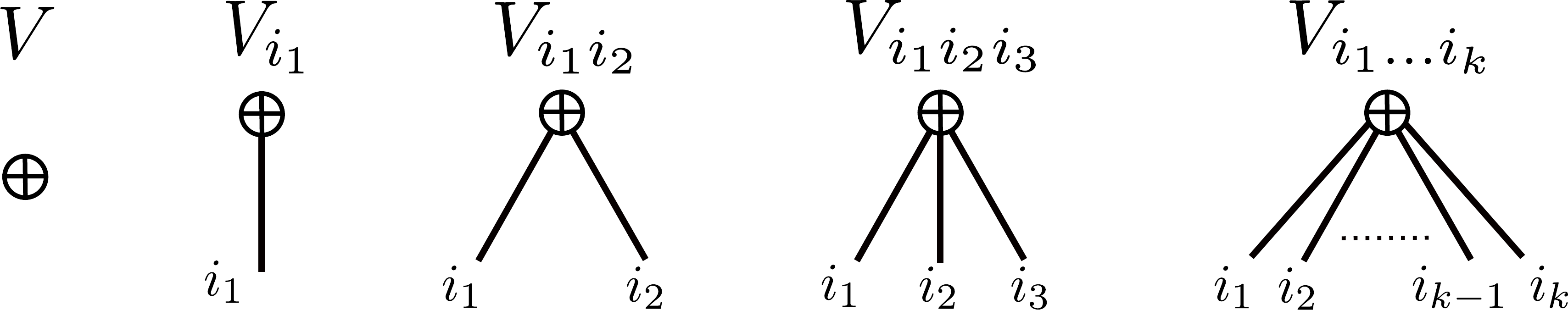}
\caption{Graphical illustration of $k$-vertex functions: The crossed circle
represents the resummation of external, unpaired particle-lines located at a
single intersection center. Whereas the outgoing lines correspond to internal
particle-lines of the backbone diagram, each carrying one of the $k$ particle
indices.}
\label{fig:vertex}
\end{figure}
Each $k$-vertex corresponds to a resummed intersection center, indicated by a
crossed circle in Fig.~\ref{fig:vertex}, and $k$ internal particle-lines.

In terms of vertex functions, the Rosenfeld functional is written as the
$0$-vertex
\begin{equation}\label{0-loop-vertex}
\beta F_{0,1} = \int V(\vec r_a)\, d^3r_a\;,
\end{equation}
whereas the Euler form of (\ref{phi-1}) translates into the product
of $2$-vertices
\begin{equation}\label{first_corr}
\begin{split}
\phi_{1,3}(\vec r_a, \vec r_b,\vec r_c)
& := \frac{1}{6}\int V^{i_1i_2}(\vec r_a) V^{i_2i_3}(\vec r_b)
V^{i_3i_1}(\vec r_c)\\
&\qquad \quad \times \rho_{i_1}\rho_{i_2}\rho_{i_3}\,
d\gamma_{i_1}d\gamma_{i_2}d\gamma_{i_3}\,.
\end{split}
\end{equation}

Vertex functions are the building blocks of FMT functionals and, in analogy to
the derivation of the 0-loop order (\ref{ros-2}), can be written as a function
of $n_v$. Setting $k=0$, Eq.~(\ref{vertex-i}) reproduces the Rosenfeld
functional:
\begin{align}
&V(\vec r_a) = \int \sum_{k=1}^\infty \frac{1}{k} 
K(\widetilde\Gamma^{(i_1\ldots i_k)}_A) \rho_{i_1}\ldots
\rho_{i_k}\,dn_v\nonumber\\
&= \int \sum_{k=1}^\infty n_\chi (n_v)^{k-1} + (k-1)\, C^{\alpha_1\alpha_2}
n_{\alpha_1} n_{\alpha_2} n_v^{k-2}\nonumber\\
&\quad  + (k-1)(k-2)\, C^{\alpha_1\alpha_2\alpha_3} n_{\alpha_1} n_{\alpha_2}
n_{\alpha_3} n_v^{k-3}\, dn_v\\
&= \int n_\chi \frac{1}{1-n_v} + C^{\alpha_1\alpha_2} n_{\alpha_1}
n_{\alpha_2} \frac{1}{(1-n_v)^2}\nonumber\\
&\quad + 2\,C^{\alpha_1\alpha_2\alpha_3} n_{\alpha_1} n_{\alpha_2} n_{\alpha_3}
\frac{1}{(1-n_v)^3}\,dn_v\;,\nonumber
\end{align}
which yields the $0$-vertex function:
\begin{align}\label{vertex-0}
V(\vec r_a) & = -n_\chi \ln{(1-n_v)} + C^{\alpha_1\alpha_2} 
n_{\alpha_1} n_{\alpha_2} \frac{1}{1-n_v}\nonumber\\
&\quad + C^{\alpha_1\alpha_2\alpha_3} n_{\alpha_1} n_{\alpha_2}
n_{\alpha_3} \frac{1}{(1-n_v)^2}\;.
\end{align}

Correspondingly, the 1-vertex is a function with a single internal
particle-line. However, as the virial expansion of the free-energy of hard
particles only depends on star-graphs, no such term will occur. Nonetheless,
the 1-vertex provides the first correction in the Mayer expansion of soft
potentials with a hard-body center. We will therefore note its form for
completeness:
\begin{align}\label{vertex-1}
V^{i_1}(\vec r_a) & = -\omega^{i_1}_\chi \ln{(1-n_v)} + C^{\alpha_1\alpha_2} 
\omega^{i_1}_{\alpha_1} n_{\alpha_2} \frac{1}{1-n_v}\nonumber\\
&\quad + C^{\alpha_1\alpha_2\alpha_3} \omega^{i_1}_{\alpha_1} n_{\alpha_2}
n_{\alpha_3} \frac{1}{(1-n_v)^2}\;.
\end{align}

The analogous calculation with two fixed particle-lines and $\rho_{i_1}$,
$\rho_{i_2}$ removed
\begin{align}
&V^{i_1i_2}(\vec r_a) = \int 
\sum_{k=2}^\infty \frac{1}{k} K(\widetilde\Gamma^{(i_1\ldots i_k)}_A)
\rho_{i_3}\ldots \rho_{i_k}\,dn_v\nonumber\\
&= \int \sum_{k=2}^\infty (k-1)\, C^{A_1A_2} \omega_{A_1}^{i_1}
\omega_{A_2}^{i_2} n_v^{k-2}\nonumber\\
&\quad  + (k-1)(k-2)\, C^{\alpha_1\alpha_2\alpha_3} \omega_{\alpha_1}^{i_1}
\omega_{\alpha_2}^{i_2} n_{\alpha_3} n_v^{k-3}\,dn_v\\
&= \int C^{A_1A_2} \omega_{A_1}^{i_1} \omega_{A_2}^{i_2}
\frac{1}{(1-n_v)^2}\nonumber\\
&\quad + 2\,C^{\alpha_1\alpha_2\alpha_3} \omega_{\alpha_1}^{i_1}
\omega_{\alpha_2}^{i_2} n_{\alpha_3}
\frac{1}{(1-n_v)^3}\,dn_v\nonumber
\end{align}
yields the $2$-vertex function
\begin{equation}\label{vertex-2}
\begin{split}
V^{i_1i_2}(\vec r_a) & = C^{A_1A_2} \omega_{A_1}^{i_1} \omega_{A_2}^{i_2}
\frac{1}{1-n_v}\\
&\quad + C^{\alpha_1\alpha_2\alpha_3} \omega_{\alpha_1}^{i_1}
\omega_{\alpha_2}^{i_2} n_{\alpha_3} \frac{1}{(1-n_v)^2}\;.
\end{split}
\end{equation}

For vertices with more than two particle-lines, the sum reduces to one term only
\begin{align}
&V^{i_1\ldots i_n}(\vec r_a) =
\int \sum_{k=n}^\infty \frac{1}{k} K(\widetilde\Gamma^{(i_1\ldots i_k)}_A)
\rho_{i_{n+1}}\ldots\rho_{i_k}\,dn_v\nonumber\\
& = \int \sum_{k=n}^\infty (n-1)!\binom{k-1}{n-1}C^{A_1A_2A_3}\nonumber\\
&\qquad \times \omega_{A_1}^{(i_1} \omega_{A_2}^{i_2}
\omega_{A_3}^{i_3} \omega_v^{i_4}\ldots \omega_v^{i_n)} n_v^{k-n}\,dn_v\\
&= \int (n-1)!\,C^{A_1A_2A_3} \nonumber\\
& \qquad  \times \omega_{A_1}^{(i_1} \omega_{A_2}^{i_2}
\omega_{A_3}^{i_3} \omega_v^{i_4}\ldots \omega_v^{i_n)}
\frac{1}{(1-n_v)^n}\,dn_v\;,\nonumber
\end{align}
whose integration yields the vertex function
\begin{align}\label{vertex-3}
&V^{i_1\ldots i_n}(\vec r_a) =(n-2)!\,C^{A_1A_2A_3} \\
&\times \omega_{A_1}^{(i_1} \omega_{A_2}^{i_2}
\omega_{A_3}^{i_3} \omega_v^{i_4}\ldots \omega_v^{i_n)}
\frac{1}{(1-n_v)^{n-1}}\quad \text{for}\; n\geq 3\;,
\nonumber
\end{align}
symmetrizised in the particle indices $i_1,\ldots, i_n$.

An important result of the resummation process is the pole structure of the
generating functions. The vertex with $k\geq 2$ particle lines has a pole at
least of order $k-1$ at packing fraction $n_v=1$. This shows that the influence
of diagrams rapidly decreases with their number of intersection centers. For the
free-energy functional of hard spheres, this explains the success of the
Rosenfeld functional. The leading correction $\phi_{1,3}$ will then be of
order $-3$ and only take affect at high densities or strong angular correlations
between particles, i.e. the solid state of the statistical system.
\section{The subtraction scheme and first order correction}
\label{subsec:functional}
Using the representation of vertex functions simplifies the derivation of new
functionals $\Phi_{g,h}$ for a given backbone diagram. However, as each order of
the expansion approximates an infinite subset of Mayer diagrams, it is not
possible to simply add its contributions. For example, the Rosenfeld functional
$\Phi_{0,1}$ approximates the third virial contribution by its contracted form,
whereas $\Phi_{1,3}$ contains its exact integral. The naive sum of both term
$\Phi_{0,1}+\Phi_{1,3}$ therefore causes a double counting of diagrams. How to
compensate such terms by subtraction will be shown in
\ref{subsubsec:first_order_correction}, followed by the resummation of Mayer
ring diagrams and subsequent comparison with the White Bear II functional in
\ref{subsubsec:white_bear_comparision}.
\subsection{The first order correction to Rosenfeld's functional}
\label{subsubsec:first_order_correction}
The splitting of intersection diagrams into subdiagrams and their subsequent
resummation is a local mapping, reflected in vertex functions that each depend
on one intersection center only. Thus the global information about the
topology of the original Mayer diagram is partially lost. Taking
Fig.~\ref{fig:butterfly} as an example, the root points of different
subdiagrams, indicated by $i,j,k$, are statistically independent, and their
Euler form has thus to vanish for concurrent intersection centers. This,
however, is in contrast to the functional (\ref{first_corr}), which is none-zero
in the limit of coincident intersection centers $B,C\to A$.

In order to obtain a physically consistent result, the degenerate contribution
for $\vec r_a=\vec r_b=\vec r_c$ has to be removed from the integral:
\begin{align}\label{F-1-3}
\beta F_{1,3} = \int &\left[ \phi_{1,3}(\vec r_a,\vec r_b,\vec
r_c) - \lim_{B,C\to A} \phi_{1,3}(\vec r_a,\vec r_b,\vec r_c)
\right]\nonumber\\
&\quad \times d\gamma_a d\gamma_b d\gamma_c
\end{align}

The subtraction of the degenerate part from the functional solves two problems:
First, it removes the contractions incompatible with the Mayer diagrams. But,
as a second effect, it also removes the free-energy contributions of lower
diagrammatic orders, that correspond to identical Mayer clusters, but at
different order of approximation. For example, the leading term of
$\phi_{1,3}(\vec r_a,\vec r_b,\vec r_c)$ is the exact third virial integral,
whereas the 3-particle contribution of the Rosenfeld functional $F_{0,1}$
contains only its approximated form as a contracted diagram. Subtracting the
contracted diagrams in (\ref{F-1-3}) removes therefore its corresponding term in
the 0-loop order of $F_{0,1}+F_{1,3}$. 

Removing unphysical terms from the free-energy is a common step in the
regularization of loop integrals in quantum field theory. Because of this
formal resemblance, we will call the current subtraction scheme the
''regularization`` of intersection diagrams.

Using the intersection algebra and the contraction rules for weight functions,
the consistency of (\ref{F-1-3}) under regularization can be shown explicitly.
Leaving out the particle densities in (\ref{first_corr}), the contraction
of the leading contribution of the 2-vertex product
\begin{align}
&\frac{1}{6}\frac{(C^{A_1A_2}\omega_{A_1}\omega_{A_2})^3}{(1-n_v)^3}
=\frac{1}{6} (C^{A_1A_2}\omega_{A_1}\omega_{A_2})^3 + \ldots \nonumber\\
&\simeq  \frac{1}{6}C^{(A_1 A_2 A_3)}\omega_{A_1} \omega_{A_2}
\omega_{A_3} 
= \frac{3}{6} C^{\chi v v } \omega_{\chi}\omega_v^2\\
& \quad + \frac{3\cdot 2}{6} C^{\alpha_1\alpha_2} \omega_{\alpha_1}
\omega_{\alpha_2}
\omega_v
+ \frac{3!}{6} C^{\alpha_1\alpha_2\alpha_3} \omega_{\alpha_1}
\omega_{\alpha_2}\omega_{\alpha_3}
\nonumber\\
&= \frac{1}{2} \omega_{\chi}\omega_v^2
+ C^{\alpha_1\alpha_2} \omega_{\alpha_1} \omega_{\alpha_2} \omega_v
+ C^{\alpha_1\alpha_2\alpha_3} \omega_{\alpha_1}
\omega_{\alpha_2}\omega_{\alpha_3}
\nonumber
\end{align}
reproduces the contracted 3-particle virial and cancels its corresponding part
in (\ref{ros-2}). In summary, the functional
\begin{equation}\label{k-funct}
F_{\text{K}}= F_{0,1} + F_{1,3}
\end{equation}
is exact up to the third virial order and approximates all triangular
and completely connected Mayer diagrams. The explicit dependence on 2-point
densities now resolves the artificial degeneracy of $F_{0,1}$ in the
orientational degrees of freedom. Furthermore, the triangular diagrams describe 
distance correlations beyond the hard-particle diameter and therefore exceed
the Percus-Yevick approximation.
\subsection{Approximations and the White Bear functional}
\label{subsubsec:white_bear_comparision}
With the given set of rules, the derivation of functional corrections of
arbitrary order becomes possible. But, as is often the case, including higher
order terms does not guarantee higher orders of precision. Possible reasons are
the divergence of series and the increasing calculational efforts to evaluate
higher order terms. Actually, the benefit of the Rosenfeld functional and the
vertex functions is their growing order in $1/(1-n_v)$, which ensures a fast
converging expansion away from its singularity. More restrictive is therefore
the second aspect, how to evaluate and minimize higher order terms, whose three
and more intersection centers define non-local functionals.

The mathematical framework necessary to evaluate such ring diagrams has been
developed by Wertheim and applied to the third virial integral for ellipsoidal
geometries at constant particle density \cite{wertheim-1, wertheim-2,
wertheim-3, wertheim-4}. In the explored range of aspect ration $L\leq 10$,
Wertheim found excellent agreement with results of computer simulations.

The same mathematical methods apply to the functional (\ref{k-funct}), with
the tensorial products of the normal vectors in (\ref{weight-functions})
represented by spherical harmonic functions and the convolute of weight
functions decoupled by a Radon transformation. The latter reduces to a Fourier
transformation in the case of coinciding intersection centers. The subtracted
part of (\ref{F-1-3}) can therefore be evaluated in the same way as the
Rosenfeld functional. Nonetheless, evaluating and minimizing the functional is
still a complicated mathematical problem, wherefore the development of efficient
approximation strategies will be an important future goal. 

One possible ansatz is the improvement of the analytical form of the
$n_v$-dependence of the functional. The previous results for the 0-loop order
and the vertex functions suggest an expansion in powers of $1/(1-n_v)$. This,
however, is a result of the chosen resummation strategy. It is important to
observe that different selections of diagrams will also yield a different
analytical structure in $n_v$. As has been discussed in \cite{korden-2}, what
marks the vertex functions as special is their possibility to combine with any
alternative resummation scheme because of the factorization of Mayer diagrams
into completely connected subdiagrams.

\begin{figure}
\includegraphics[width=2.2cm,angle=-90]{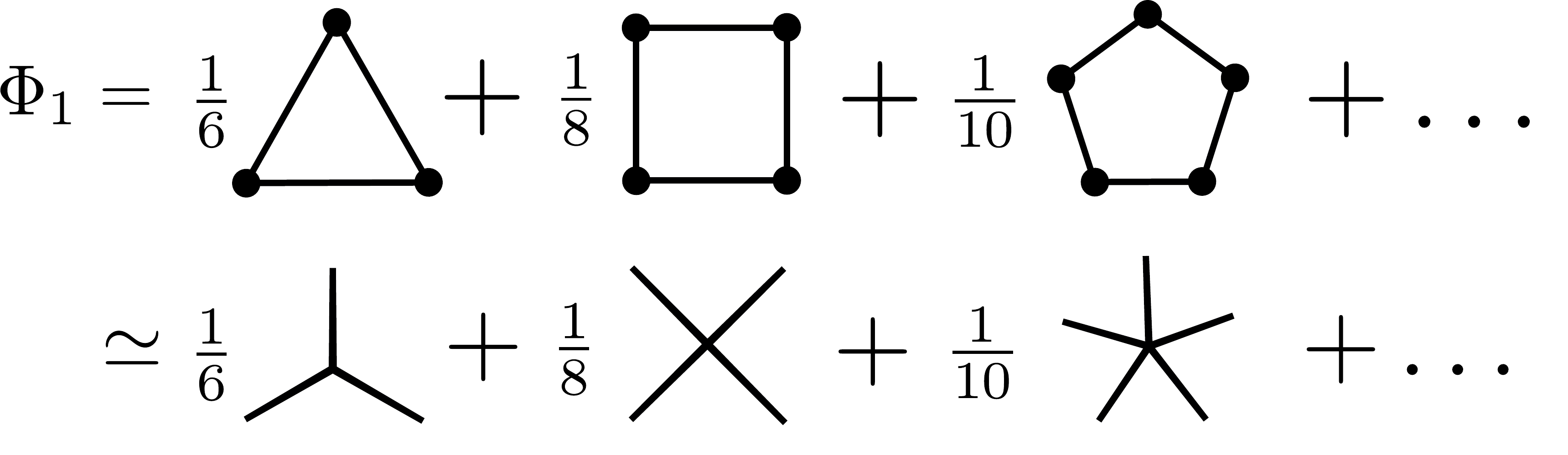}
\caption{$\Phi_1$ is the generating functional for all 1-loop Mayer diagrams,
whose contraction provides a first order approximation in $n_v$ to the
Rosenfeld functional.}
\label{fig:wb-approx}
\end{figure}

In order to improve the $n_v$-dependence of the 0-loop order, it is necessary
to go beyond the starfish graphs. The first choice is therefore the set of
1-loop diagrams
\begin{equation}
\Phi_1 = \sum_{h=3}^\infty \Phi_{1,h}\;.
\end{equation}
As a further approximation we restrict the functionals $\Phi_{g,h}$ to their
backbone diagrams, which results in the series shown in the first line of
Fig.~\ref{fig:wb-approx}.

The functional $\Phi_1$ for Mayer ring-diagrams has already been derived in
\cite{korden-2} as the generating function of Mayer bonds $f_{i_1i_2}$, written
in the matrix notation
\begin{equation}
M_{i_1i_2} := C^{A_1A_2}\omega_{A_1}^{i_1}\omega_{A_2}^{i_2}\rho_{i_1}\;.
\end{equation}
Together with the symmetry factor $\sigma(k) = 1/(2k)$ for a ring of $k$
particles, the sum can be rewritten in closed form
\begin{align}\label{1-loop-sum}
\Phi_1(M) & = \sum_{k=3}^\infty \;\frac{1}{2k}\;
M_{i_1i_2}M_{i_2i_3}\ldots M_{i_ki_1}\\
& = -\frac{1}{2}\ln{(1-M)} - \frac{1}{2}M_{i_1i_1} -
\frac{1}{4}M_{i_1i_2}M_{i_2i_1}\nonumber
\end{align}
with a logarithmic singularity at $M=1$. It is tempting to assume that this
divergency corrects the pole $n_v=1$ of highest packing fraction of the 0-loop
functional to a physically realistic value that depends on the geometry of the
particles. However, minimizing $\Phi_1$ will be even more ambitious than that of
$F_\text{K}$. 

As we are only interested in the $n_v$-corrections of the Rosenfeld functional,
it is sufficient to contract (\ref{1-loop-sum}) to one intersection center.
Using the notation 
\begin{equation}
\Gamma_{i_1i_2\ldots i_k} :=
<\Gamma_{i_1i_2}\Gamma_{i_2i_3}\ldots \Gamma_{i_ki_1}>
\end{equation}
to indicate the contraction of a Mayer diagram, the logarithmic part of
(\ref{1-loop-sum}) can be expanded in orders of $n_v$:
\begin{align}
&-\frac{1}{2}<\ln{(1-M)}> = \sum_{k=1}^\infty \;\frac{1}{2k}<M^k>\nonumber\\
&= \int \sum_{k=1}^\infty\;\frac{1}{2k}\;K(\frac{1}{k}
\Gamma_{i_1\ldots i_k})\rho^k \,dn_v \nonumber\\
&= \int \sum_{k=1}^\infty\; \frac{1}{2k}C^{(A_1\ldots A_k)}n_{A_1}\ldots
n_{A_k}\int_0^1 t^{k-1}d(tn_v)\label{log-calc}\\
&=\frac{1}{2}\int \frac{n_\chi}{1-tn_v} + C^{\alpha_1\alpha_2} n_{\alpha_1}
n_{\alpha_2} \frac{t}{(1-tn_v)^2} \nonumber\\
&\quad + 2\, C^{\alpha_1\alpha_2\alpha_3} n_{\alpha_1} n_{\alpha_2} n_{\alpha_3}
\frac{t^2}{(1-tn_v)^3}\,d(tn_v)\;,\nonumber
\end{align}
where the integration over the scaling parameter $\rho\to t\rho$ absorbs one
factor of $1/k$. Evaluating the integral yields the logarithmic contribution
\begin{align}
&-\frac{1}{2}<\ln{(1-M)}> = -\frac{1}{2} n_\chi \ln{(1-n_v)}\label{wb-k}\\
&+ C^{\alpha_1\alpha_2} n_{\alpha_1} n_{\alpha_2}
\frac{1}{4n_v(1-n_v)} [2(1-n_v)\ln{(1-n_v)} + 2n_v] \nonumber\\
&- C^{\alpha_1\alpha_2\alpha_3} n_{\alpha_1} n_{\alpha_2} n_{\alpha_3}
\nonumber \\
&\quad \times \frac{1}{2n_v^2(1-n_v)^2}[2(1-n_v)^2\ln{(1-n_v)} + 2n_v - 3n_v^2]
\;.\nonumber
\end{align}

The remaining two terms of (\ref{1-loop-sum}) are the traces of $M$ and $M^2$,
which are functions of one intersection center only. Thus, using the subtraction
scheme introduced in the last paragraph, both contributions would vanish after
regularizing the 1-loop diagrams. Unfortunately, this mechanism does not apply
for the current approximation and a better understanding of these two terms is
necessary.

Deriving the trace of the contracted form of $M$ 
\begin{equation}
\begin{split}
<M_{i_1i_1}> & = <C^{A_1A_2}\omega_{A_1}^{i_1}\omega_{A_2}^{i_2}>\rho_{i_1}\\
&= <C^A\omega_A^{i_1}>\rho_{i_1} = C^\chi \omega_\chi^{i_1}\rho_{i_1} = n_\chi
\end{split}
\end{equation}
yields the weight density of the Euler characteristic. It therefore removes in 
(\ref{1-loop-sum}) the case of only one intersecting particle. This geometric
interpretation is consistent with Fig.~\ref{fig:wb-approx}, where at least three
particles have to interact at one intersection point. Generalizing this
observation to the calculation (\ref{log-calc}), we have to subtract the
contributions of $K(\Sigma\cap D^{k-1})$ from its sum. The derivation is
analogous to the calculation of the 1-vertex (\ref{vertex-1}), only with the
symmetry factor replaced by $\sigma(1)=1/2$ and the sum restricted to
$\Sigma\cap D^{k-1}$:
\begin{equation}\label{first-delete}
-\frac{1}{2}\int \sum_{k=1}^\infty C^{\chi v \ldots v}n_\chi n_v^{k-1}\,dn_v
= \frac{1}{2} n_\chi \ln{(1-n_v)}
\end{equation}

The calculation for the second term $M^2$ is similar but allows
two possible contractions:
\begin{align}
<M_{i_1i_2} & M_{i_2i_1}>\nonumber\\
& = <C^{A_1A_2}\omega_{A_1}^{i_1}\omega_{A_2}^{i_2}
C^{A_2A_1}\omega_{A_2}^{i_2}\omega_{A_1}^{i_1}> \rho_{i_1}\rho_{i_2}\nonumber\\
&= C^{\alpha_1\alpha_2}n_{\alpha_1}n_{\alpha_2}\nonumber\\
<M_{i_1i_3}& M_{i_3i_2}>\rho_{i_3}\\
& = <C^{A_1A_3}\omega_{A_1}^{i_1}\omega_{A_3}^{i_3}
C^{A_3A_2}\omega_{A_3}^{i_3}\omega_{A_2}^{i_2}> \rho_{i_1}\rho_{i_2}\rho_{i_3}
\nonumber\\
&= C^{\alpha_1\alpha_3\alpha_2}n_{\alpha_1}n_{\alpha_3}n_{\alpha_2}\;,\nonumber
\end{align}
corresponding to intersections of $\Sigma^2$ and $\Sigma^3$. But as both
contributions derive from the same order in $M$, they necessarily have to follow
from the same $C$-matrix. With the symmetry factor $\sigma(2)=1/4$, the terms to
be removed from (\ref{wb-k}) are the Euler forms of $K(\Sigma^2\cap D^{k-1})$
and $K(\Sigma^3\cap D^{k-1})$ multiplied by the same order of $D^{k-1}$ as in
(\ref{first-delete}). The calculation therefore parallels (\ref{vertex-2}) with
$n_\chi$ removed:
\begin{equation}\label{second-delete}
\begin{split}
-\frac{1}{4}V^{i_1i_2}& \rho_{i_1}\rho_{i_2}n_v
=-\frac{1}{4} C^{\alpha_1\alpha_2}n_{\alpha_1}n_{\alpha_2}\frac{n_v}{1-n_v}\\
& -\frac{1}{4}C^{\alpha_1\alpha_2\alpha_3}n_{\alpha_1}
n_{\alpha_2}n_{\alpha_3}\frac{n_v}{(1-n_v)^2}\;.
\end{split}
\end{equation}

The new functional is the sum of the three contributions (\ref{wb-k}),
(\ref{first-delete}), (\ref{second-delete}) and the 0-loop order (\ref{ros-2}):
\begin{equation}\label{mark-2}
\begin{split}
\Phi &=-n_\chi\ln{(1-n_v)}(1+\phi^{(1)})\\
& + C^{\alpha_1\alpha_2} n_{\alpha_1}n_{\alpha_2}\frac{1}{1-n_v}
(1+\phi^{(2)})\\
& + C^{\alpha_1\alpha_2\alpha_3}
n_{\alpha_1}n_{\alpha_2}n_{\alpha_3}\frac{1}{(1-n_v)^2} 
(1+\phi^{(3)})
\end{split}
\end{equation}
with the three correction terms:
\begin{align}\label{corr-k}
\phi_{\text{K}}^{(1)} &= 0\nonumber\\
\phi_{\text{K}}^{(2)} &= \frac{1}{4n_v} (2(1-n_v)\ln{(1-n_v)}+2n_v -n_v^2)\\
\phi_{\text{K}}^{(3)} &= \frac{-1}{2n_v^2}(2(1-n_v)^2\ln{(1-n_v)}+2n_v-3n_v^2
+\frac{1}{2} n_v^3)\;.\nonumber
\end{align}

In this form, the result can be compared to the White Bear II functional,
introduced in \cite{white-bear-1, white-bear-2}, which combines the
Boublik-Mansoori-Carnahan-Starling-Leland equation of state \cite{mcsl} with the
structure of the free energy functional, determined by the scaled-particle
differential equation \cite{rosenfeld-structure}. It is therefore not a purely
geometrically motivated approach as Rosenfeld's FMT, but takes into account the
numerically derived virial coefficients up to the eighth's order, combined in a
generating function \cite{cs}.

The WBII functional is independent of positional correlations and therefore of
the same structure as (\ref{mark-2}) with the corresponding correction terms:
\begin{align}
\phi_{\text{WB}}^{(1)}& = 0\nonumber\\
\phi_{\text{WB}}^{(2)}& = \frac{1}{3n_v} [2(1-n_v)\ln{(1-n_v)}+2n_v-n_v^2]
\label{corr-wb}\\
\phi_{\text{WB}}^{(3)} &= 
\frac{-1}{3n_v^2} [2(1-n_v)^2\ln{(1-n_v)} +2n_v-3n_v^2+2n_v^3])\;.\nonumber
\end{align}

The terms $\phi_\text{K}$, $\phi_\text{WB}$ are similar, deviating only in their
numerical prefactors and the $n_v^3$-term of $\phi^{(3)}$. Comparing the curve
shapes of $\phi^{(2)}_\text{K}$ and $\phi^{(2)}_\text{WB}$, we find excellent
agreement despite this property. Whereas the different $n_v^3$-terms of
$\phi^{(3)}$ cause a significant change in the curve's curvature. Nonetheless,
it is remarkable that the analytical terms of the infinite sum of Mayer diagrams
(\ref{wb-k}) are in good agreement with their analytical counterparts in the
White Bear II functional.
\section{Discussion and Conclusion}\label{sec:conclusion}
The current article has shifted the previous perspective of \cite{korden-2} from
differential geometry to the algebraic rules of the Euler form. It
systematically generalizes the FMT functional from Rosenfeld's 0-loop order to
any number of intersection centers and develops several new mathematical tools
for the efficient manipulation of weight functions.

It has been shown that Mayer's star graphs have a uniquely defined dual
representation in intersection diagrams. These allow an intuitive picture of
the Euler form and its decomposition into weight functions. Using this
graphical description, we developed the contraction method as an approximation
of the underlying Mayer diagrams. 

Removing the external particle-lines from an intersection diagram defines its
backbone graph, which corresponds to a unique contribution $\Phi_{g,h}$ in the
expansion of the free energy functional. The successive attachment of external
particle-lines to its intersection centers provides a resummation process that
significantly improves the original virial expansion in particle densities. The
resulting vertex functions then replace the Mayer functions as the building
blocks of the FMT functional.

Resummation is an essential step in the derivation of the functional as it
generates the pole structure $1/(1-n_v)$ of the packing fraction and yields a
generic convergence criterion for the expansion in intersection centers.
However, its factorization in vertex functions could not be proven in general.
The problem relies in the symmetry factors of the infinite sum of Mayer
diagrams. For the exemplary case of triangular graphs it could be shown that
such a splitting indeed exists as the free-energy prefactor for all such
diagrams agrees.

In the current case of triangular diagrams, we first determined the invariance
groups under labeling which define four equivalence classes. Furthermore, taking
into account the contraction and polynomial multiplicities of their individual
vertex contributions, it could be shown that each triangular diagram contributes
the same numerical prefactor. This allowed the simple factorization into fully
contracted subdiagrams for any triangular graph. For general diagrams, however,
we were only able to show that the prefactors for a given backbone diagram
transform under the same automorphism group as the diagram itself. Nevertheless,
this is enough to deduce that the resummed free energy functional for any Mayer
ring-diagram likewise factorizes, with a common prefactor that only depends on
its number of intersection centers. But for more general functional
contributions of two and more loops, it would be an important step in
our understanding to obtain a general expression for this structure.

The resummation of diagrams also involved the regularization of the functional.
This process removes those intersection terms that are either incompatible with
the Mayer diagrams or contributions of lower order functionals which are
now replaced by terms of lesser approximation. Most terms of the subtraction
scheme are of zero measure and thus can be ignored completely. A practical
application is the summation over all completely contracted Mayer ring-diagrams.
The analytical form reproduces the White Bear II functional to good accuracy.
However, it is the subtraction of the irregular parts of first and second order
that cause the deviation in $n_v^3$ that does not match with the White Bear
result. If this discrepancy could be clarified it would be possible to derive
even higher order corrections and exceed the current precision of numerically
obtained functionals. 

The White Bear II functional provides only a small correction to the bulk
properties of the Rosenfeld functional. This suggests that the improvement of
the analytical structure in $n_v$ will be less important than the inclusion of
terms which resolve the orientational degeneration of the 0-loop order and to go
beyond the Percus-Yevick approximation. Both deficits are resolved by the
three-center term derived in this article. It will therefore be an important
next step to find efficient numerical methods to minimize $F_\text{K}$.

The new objects entering this functional are the 2-point densities. Whereas the
1-point functions depend on a single vector field moving over the surface of a
particle, the 2-point functions determine the correlation between vector fields
at two different particle points. Thus, 1-point functions reproduce the
''classical`` curvature depend information of the particle's geometry, whereas
$k$-point functions define a completely new mathematical class of geometric
invariant quantities. Fundamental measure theory introduces therefore new
mathematical tools that not only give access to important physical problems of
many-particle systems but also might provide new answers to mathematical
questions that cannot be covered by single vector fields alone. $k$-point
functions interpolate between the manageable but approximate description of
geometry as a tangential space and non-local geometric properties as, e.g., the
maximal packing density of particles in an embedding space.

The derivation of a FMT functional starting from an infinite class of Mayer
graphs, then translated to intersection diagrams, and finally resummed into
vertex functions is an inefficient approach. Given the diagrams' systematic and
our first experience with ring graphs, we expect the existence of a simpler
formulation that closely resembles a field theory with the vertex functions as
its variabels. This would completely replace the dependence on Mayer diagrams
and their symmetry factors and provide a better understanding of the nature of
FMT.
\section{Acknowledgment}
Matthias Schmidt is kindly acknowledged for helpful discussions and his infinite
patience while preparing this work. The author also wishes to thank Andr\'e
Bardow and Kai Leonhard for supporting this work, as well as Annett Schwarz and
Christian Jens for carefully proofreading the manuscript.
This work was performed as part of the Cluster of Excellence "Tailor-Made Fuels
from Biomass", funded by the Excellence Initiative of the German federal and
state governments.
%
%
%
%
\appendix
\section{}\label{sub:appendix_a}
In the following we will determine the automorphism groups and their
characteristic parameters (\ref{sym_mayer}), (\ref{sym-i}), and (\ref{kappa})
for the triangular Mayer diagrams $\Gamma(n_1,n_2,n_3)$ of the type shown in
Fig.~\ref{fig:butterfly}. To the best of the authors knowledge, no general
classification or systematic construction of these groups is known. We will
therefore first derive the groups of the first seven diagrams and then deduce
their generalization to all further cases.

To put the formulation on a more formal level, consider any star diagram
$\Gamma$ and define a representation $\lambda$ of $\Gamma$ by its labeling in
Mayer's $f$ functions:
\begin{equation}
\lambda\; : \; \Gamma \to \text{prod}(f_{ij})\;.
\end{equation}
By definition, any product of $f$ functions is only uniquely defined up to their
ordering
\begin{equation}
\begin{split}
\pi := \{f_{ij} = f_{ji} , \; & f_{ij}f_{kl} = f_{kl}f_{ij}\}\\[0.5em]
\pi \lambda(\Gamma) &  = \lambda(\Gamma)\;.
\end{split}
\end{equation}

As any labeling of the nodes of a Mayer diagram is admissible, there exists
$n!$ possible representations for an $n$-particle diagram, generated by
operating with the symmetric group $S_n$ on any representation $\lambda$:
\begin{equation}
g\in S_n \;: \; g\lambda = \lambda'\;.
\end{equation}
For a discussion of the symmetric group see e.g. \cite{hamermesh}. However, not
all elements of $S_n$ generate a new representation. If the operation of $g$ can
be undone by a permutation $\pi$, it leaves the labeling invariant, defining a
subgroup of $S_n$:
\begin{equation}
\text{Aut}(\Gamma) = \{ g\in S_n \; | \; g\lambda(\Gamma) = \pi
\lambda(\Gamma)\}\;,
\end{equation}
so that $\pi^{-1} \circ g = \text{id}$. Clearly, the identity $e\in S_n$ is
element of $\text{Aut}(\Gamma)$, and if $g_1, g_2 \in \text{Aut}(\Gamma)$ so is
$g_1^{-1}, g_2^{-1}$ and any product of them. 

To simplify the notation, let us replace the $f$ functions by square brackets:
\begin{equation}
f_{ij} \to [ij]\;,
\end{equation}
on which the permutation symbols $(1,2,3,\ldots)$ of $S_n$ operate by
cyclic permutation of particle indices. 

As an example, consider the labeling of the diagram $\Gamma(1,1,0)$, as shown in
Fig.~\ref{fig:mayer_anhang} with its $\mathbb{Z}_2$-symmetry $2\leftrightarrow
3$, $4\leftrightarrow 5$:
\begin{figure}
\includegraphics[width=3.5cm,angle=-90]{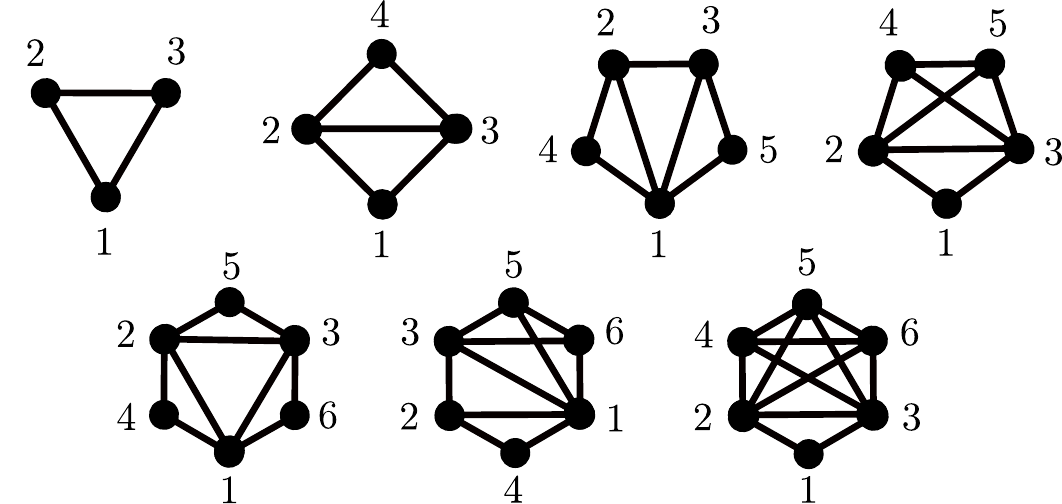}
\caption{The first Mayer diagrams of up to 6 particles with triangular
substructure. $(0,0,0)$, $(1,0,0)$, $(1,1,0)$, $(2,0,0)$, $(1,1,1)$, $(2,1,0)$,
and $(3,0,0)$ with the inner triangle defined by the numbers $1\to 2 \to 3$.}
\label{fig:mayer_anhang}
\end{figure}
\begin{equation}
\begin{split}
\lambda(\Gamma_{1,1,0}) & = f_{12}f_{13}f_{14}f_{15}f_{23}f_{24}f_{35} \\ 
&= [12][13][14][15][23][24][35]\;.
\end{split}
\end{equation}

The axial symmetry of the diagram only allows the identity and one further
element as the automorphism group
\begin{equation}
\begin{split}
(1)(23)(45) & \Bigl([12][13][14][15][23][24][35]\Bigr)\\
&\qquad  = [13][12][15][14][32][35][24]
\end{split}
\end{equation}
so that the order of $\text{Aut}(\Gamma(1,1,0))$ is 2.

The graphical representation of the diagrams, shown in
Fig.~\ref{fig:mayer_anhang}, suggests a relation to point-groups. This is of
course only true for the simple graphs under consideration, but simplifies the
construction of the automorphism groups considerably. For up to six particles,
the star diagrams and group elements are as follows:
\begin{align}
&(0,0,0): S_3\label{A1}\\
& (1)(2)(3), (1)(23), (2)(13), (3)(12), (123), (132)\nonumber\\[0.5em]
&(1,0,0): \mathbb{Z}_2\times S_2\\
&(1)(2)(3)(4), (1)(4)(23), (2)(3)(14), (14)(23)\nonumber\\[0.5em]
&(1,1,0):  E_1^3\times \mathbb{Z}_2\\
&(1)(2)(3)(4)(5), (1)(23)(45)\nonumber\\[0.5em]
&(2,0,0):  E_1\times \mathbb{Z}_2\times S_2\\
&(1)(2)(3)(4)(5), (1)(2)(3)(45), (1)(4)(5)(23),\nonumber\\
&(1)(23)(45)\nonumber\\[0.5em]
&(1,1,1):  G_6 \label{111}\\ 
&(1)(2)(3)(4)(5)(6), (1)(5)(23)(46), (2)(6)(13)(45),\nonumber\\
&(3)(4)(12)(56), (123)(456), (132)(654)\nonumber\\[0.5em]
&(2,1,0):  E_1^4\times S_2\\
&(1)(2)(3)(4)(5)(6), (1)(2)(3)(4)(56)\nonumber\\[0.5em]
&(3,0,0):  E_1\times S_2\times S_3\label{A7}\\
&(1)(2)(3)(4)(5)(6), (1)(2)(3)(4)(56), (1)(2)(3)(5)(46),\nonumber\\
&(1)(2)(3)(6)(45), (1)(2)(3)(456), (1)(2)(3)(465), \nonumber\\
&(1)(23)(4)(5)(6), (1)(23)(4)(56), (1)(23)(5)(46),\nonumber\\
&(1)(23)(6)(45), (1)(23)(456), (1)(23)(465)\;,\nonumber
\end{align}
where $E_1$ is the identity element. The diagrams and groups are also listed in
Tab.~\ref{tab:group_small}.

\begin{table}
\caption{\label{tab:group_small} The triangular Mayer diagrams for $n\leq 6$
particles are listed with their automorphism groups and corresponding
characteristic numbers: the symmetry factors $\sigma$, $\widetilde \sigma$,
the polynomial $p$ and contraction multiplicities $m$, and their resulting
prefactors $\widetilde\kappa$ for the free energy functional.}
\begin{ruledtabular}
\begin{tabular}{c|c|c|c|c|c|c}
$\Gamma$ & $\text{Aut}(\Gamma)$ & $\sigma$ & $\widetilde\sigma$ &
$p$ & $m$ &$\widetilde \kappa$ \\\hline
$(0,0,0)$ & $S_3$ & 1 & 1/6  & 1 & 1 & 1/6\\
$(1,0,0)$ & $\mathbb{Z}_2\times S_2$ & 6 & 1/4 & 3 & 2 &1/6\\
$(1,1,0)$ & $E_1^3\times \mathbb{Z}_2$ & 60 & 1/2 & 3 & 1 & 1/6\\
$(2,0,0)$ & $E_1\times \mathbb{Z}_2\times S_2$ & 30 & 1/2 & 3 & 1 & 1/6\\
$(1,1,1)$ & $G_6$ & 120 & 1/6 & 1 & 1 & 1/6\\
$(2,1,0)$ & $E_1^4\times S_2$ & 360 & 1 &  6 & 1 & 1/6\\
$(3,0,0)$ & $E_1\times S_2\times S_3$ & 60 & 1/2 & 3 & 1 & 1/6
\end{tabular}
\end{ruledtabular}
\end{table}
These groups have already been identified by Riddell and published by Uhlenbeck
and Ford \cite{uhlenbeck-ford-1}. Based on Polya's counting theorem
\cite{polya-a, polya-b}, they also developed a counting formula \cite{riddell,
uhlenbeck-ford-2}, which determines the number of independent labelings of all
Mayer diagrams for a given number of nodes and labels. But a corresponding
formula for individual diagrams is still unknown. 

For the current case of triangular Mayer diagrams $\Gamma(n_1,n_2,n_3)$, the
automorphism groups can be derived from the seven cases (\ref{A1}-\ref{A7}). By
comparing their groups with the diagrams of Fig.~\ref{fig:mayer_anhang}, one
observes that the three attached completely connected subdiagrams contribute the
symmetric group $S_{n_1}\times S_{n_2}\times S_{n_3}$. For $n_1\neq n_2\neq
n_3$, the resulting automorphism group therefore is $E_1^3\times S_{n_1}\times
S_{n_2}\times S_{n_3}$. For $n_1= n_2\neq n_3$, the diagram has an additional
axial symmetry $\mathbb{Z}_2$, exchanging the two subgroups $S_{n_1}\times
S_{n_2}$. And for $n_1= n_2 = n_3$, the symmetry of the backbone diagram extends
to the dihedral group $D_6$, whose semi-direct product with $S_n^3$ is denoted
as $G_{3n+3}$.

As an example, how to generalize the above seven cases, consider
$\Gamma(1,1,1)$ and its extension to $\Gamma(n,n,n)$. In the corresponding
diagram of Fig.~\ref{fig:mayer_anhang}, one replaces the particles $4,5,6$ by a
completely connected subdiagram of $n$ particles. The dihedral symmetry of the
backbone graph remains unchanged under this operation, so that the group
elements (\ref{111}) can be adjusted by the formal replacement of the particle
indices $4,5,6 \to A,B,C$ with $A,B,C \in S_n$:
\begin{align}
& (n,n,n) : G_{3n+3}:\\
& (1)(2)(3)(A)(B)(C),\; (1)(23)(B)(AC), (2)(13)(C)(AB),\nonumber\\ 
&(3)(12)(A)(BC), (123)(ABC),\; (132)(CBA)\;.\nonumber
\end{align}
The resulting automorphism group is thus the semi-direct product $D_6 \rtimes
S_n^3$.

Comparing these invariance groups to Tab.~\ref{tab:group_small}, the only
diagram that drops out of this classification is $\Gamma(1,0,0)$. Instead of
$\mathbb{Z}_2 \times E_1^2$, as expected, its automorphism group is
$\mathbb{Z}_2 \times S_2$. The reason for this larger group is an additional
$\mathbb{Z}_2$ symmetry of its diagram, exchanging the two subtriangles $1-2-3$
and $2-3-4$, using the numbering of Fig.~\ref{fig:mayer_anhang}. As has already
been observed in Section \ref{subsec:inter_diagrams}, the contraction of the
completely connected subdiagrams is therefore not unique and we have to count
the contraction multiplicity $m=2$ for $(1,0,0)$ and $m=1$ for all other
triangular diagrams.
\begin{table}
\caption{\label{tab:group_all} Complete list of the triangular Mayer diagrams
and their automorphism groups, their intersection symmetry numbers $\widetilde
\sigma$, polynomial multiplicities $p$, and free energy coefficients $\widetilde
\kappa$.}
\begin{ruledtabular}
\begin{tabular}{c|c|c|c|c}
$\Gamma(n_1,n_2,n_3)$ & $\text{Aut}(\Gamma)$ & $\widetilde\sigma$ &
$p$ &$\widetilde \kappa$ \\\hline
$n_1\neq n_2\neq n_3 $ & $E_1^3\times S_{n_1}\times S_{n_2}\times
S_{n_3}$& 1 & 6 & 1/6\\
$n_1 = n_2\neq n_3$ & $E_1\times \mathbb{Z}_2\times S_{n_1}^2\times
S_{n_3} $& 1/2 & 3 & 1/6 \\
$n_1= n_2= n_3$ & $G_{3n+3}$ & 1/6 & 1 & 1/6 \\
$n_1= n_2=0, n_3=1$ & $\mathbb{Z}_2 \times S_2$ & 1/4 & 3 & 1/6
\end{tabular}
\end{ruledtabular}
\end{table}
Adding the exceptional case $(1,0,0)$ to the previous list of triangular
diagrams, all automorphism groups have been identified and are summarized in
Tab.~\ref{tab:group_all}. 

The irreducible set of labeled diagrams can now be obtained by operating with
the subgroup 
\begin{equation}
S_n/\text{Aut}(\Gamma_n)
\end{equation}
on one representative element $\lambda(\Gamma_n)$. The number of differently
labeled diagrams is therefore the order of this group, conventionally noted by
the symmetry factor (\ref{sym_mayer}) of the virial integral
(\ref{virial-integral}). For the contracted dual diagram $\widetilde \Gamma$,
the corresponding symmetry factor has been defined in (\ref{sym-i}), which for
triangular diagrams simplifies to
\begin{equation}
\widetilde\sigma(\Gamma(n_1,n_2,n_3))=\left|
\frac{S_{n_1}\times S_{n_2}\times
S_{n_3}}{\text{Aut}(\Gamma(n_1,n_2,n_3))}\right|\;.
\end{equation}

As has been discussed in section \ref{subsec:vertex}, splitting the virial
integral into vertex functions (\ref{phi-1}), (\ref{phi-2}), (\ref{phi-3}) is
not uniquely defined but yields a multiple counting of virial diagrams
$\Gamma(n_1,n_2,n_3)$ by permutation of its polynomial factors into
$\Gamma(n_1)^{i_1i_2} \Gamma(n_2)^{i_2i_3} \Gamma(n_3)^{i_3i_1}$. For $k$
identical values of the triplet $n_1,n_2,n_3$, the polynomial multiplicity is
\begin{equation}
p = \frac{3!}{k!}\;.
\end{equation}

The overall symmetry factor $\widetilde \kappa$ for the decoupled virial
integral is therefore the product
\begin{equation}
\widetilde\kappa= \frac{m}{p}\,\widetilde\sigma\;,
\end{equation}
whose numerical values are listed in Tab.~\ref{tab:group_small} and
TAB.~\ref{tab:group_all}. The central result of the current discussion is
the free-energy prefactor, which for all triangular diagrams has the same 
numerical value
\begin{equation}\label{1/6}
\widetilde \kappa = \frac{1}{6}\;,
\end{equation}
as necessary for the decoupling and resummation of the intersection diagrams
into vertex functions.

For more complex classes of Mayer diagrams, the derivation of the automorphic
groups is similar. However, apart from the exceptional cases, it is simpler to
replace the Mayer diagrams by ''weighted intersection graphs``. They follow
from the maximally contracted intersection diagrams by noting the number of
unpaired particle lines at each vertex center and the subsequent removal of
those lines from the diagram, as is shown in Fig.~\ref{fig:butterfly}. For
planar graphs, as the triangular example, the invariance group of the Mayer
diagram derives from the semi-direct product of the symmetric groups of the
vertices with the invariance group of the weighted intersection diagram. As the
reduced diagram can be drawn in the plane, the latter is a discrete subgroup of
$\text{O}(2, \mathbb{R})$. For more general cases the weighted intersection
diagram can always be embedded into a Riemannian surface $T_g$ of minimal genus
$g$ \cite{diestel}. The invariance group of the weighted intersection diagrams
is therefore either a discrete subgroup of $\text{O}(3, \mathbb{R})$ for the
sphere $g=0$ or a subgroup of $\text{Sp}(2g,n)$ for a diagram embedded into a
torus of genus $g$. This dependence of the automorphism group on the topology
of the graph might give a first explanation why Polya's counting theorem is
applicable only for diagrams of individual classes.
\bibliography{Rosenfeld_Loop}
\end{document}